\documentclass[aps,prd,noshowpacs,nofootinbib]{revtex4}
\usepackage{dcolumn}
\usepackage{amsmath,amssymb,setspace,graphicx,subfigure,epsfig}
\usepackage{amsfonts}
\usepackage{latexsym}
\usepackage{epsfig}
\usepackage{color}
\usepackage{nicefrac}

\newcommand{\dis}[1]{\begin{equation}\begin{split}#1\end{split}\end{equation}}

\newcommand{\be}{\begin{equation}}
\newcommand{\ee}{\end{equation}}
\newcommand{\eq}[1]{Eq.~(\ref{#1})}
\newcommand{\bfrac}[2]{{\left(\frac{#1}{#2} \right)  }}

\newcommand{\etal}{{\it et al}}

\newcommand{\Mp}{M_P}

\newcommand{\p}{\partial}
\newcommand{\vp}{\varphi}
\newcommand{\vps}{\varphi_*}
\newcommand{\chis}{\chi_*}

\newcommand{\epstar}{\epsilon_\vp^*}
\newcommand{\ep}{\epsilon_\vp}

\newcommand{\ecstar}{\epsilon_\chi^*}
\newcommand{\ec}{\epsilon_\chi}

\newcommand{\fnl}{f_{\rm NL}}

\newcommand{\calA}{{\cal A}}

\newcommand{\calP}{{\cal P}}

\newcommand{\calG}{{\cal G}}

\def\ses{\sin^2\theta^e}
\def\sis{\sin^2\theta^*}
\def\etap{\eta_{\varphi\varphi}}
\def\etac{\eta_{\chi\chi}}

\def\bea{\begin{eqnarray}}
\def\eea{\end{eqnarray}}

\def\bkone{{\bf k_1}}
\def\bktwo{{\bf k_2}}

\def\picube{(2\pi)^3}
\def\la{\langle}
\def\ra{\rangle}
\newcommand{\sdelta}[1]{\!\delta^{\,3}(\mathbf{#1})}

\def\tr{\tilde{r}}

\def\Tr{T_{{\rm reh}}}

\begin{document}
\title{Review of local non-Gaussianity from multi-field inflation}
\author{Christian T. Byrnes}
\email{Byrnes@physik.uni-bielefeld.de}
\affiliation{Fakult{\"a}t f{\"u}r Physik, Universit{\"a}t Bielefeld,
Postfach 100131, 33501 Bielefeld, Germany}
\author{Ki-Young Choi}
\email{kiyoung.choi@pusan.ac.kr}
\affiliation{Department of Physics, Pusan National University, Busan,
609-735, Korea}


\hfill BI-TP 2010/02

\hfill PNUTP-10-A07


\begin{abstract}
We review models which generate a large non-Gaussianity of the local form. We first briefly consider three models which generate the non-Gaussianity either at or after the end of inflation; the curvaton scenario, modulated (p)reheating and an inhomogeneous end of inflation. We then focus on ways of generating the non-Gaussianity during inflation. We derive general conditions which a product or sum separable potential must satisfy in order to generate a large local bispectrum during slow-roll inflation. As an application we consider two-field hybrid inflation. We then derive a formalism not based on slow roll which can be applied to models in which the slow-roll parameters become large before inflation ends. An exactly soluble two-field model is given in which this happens. Finally we also consider further non-Gaussian observables; a scale dependence of $\fnl$ and the trispectrum.
\end{abstract}

\maketitle


\section{Introduction}\label{sec:introduction}

There are many models of the universe which can predict a large non-Gaussianity.
However the predicted amplitude and the shape of the non-Gaussianity 
are different among different classes of models.
One category is those which generate the non-Gaussianity due to non-trivial classical 
dynamics on superhorizon scales.
These models predict the shape of the bispectrum to be of the so called `local type',
which can be expressed as an expansion of the Bardeen potential~\cite{Komatsu:2001rj}
\dis{\label{Phifnl}
\Phi({\bf x})=\Phi_L({\bf x}) + \fnl( \Phi_L^2({\bf x})-\langle\Phi_L^2({\bf x})\rangle ),
}
where $\Phi$ is the curvature perturbation on a Newtonian slice and $\Phi_L$
is its linear and Gaussian part. $\langle\Phi_L^2({\bf x})\rangle$ denotes the
ensemble average in a statistically homogeneous distribution.
The current limit on the local type of the non-linearity parameter $\fnl$ 
from seven years of WMAP data~\cite{Komatsu:2010fb} is 
$-10<\fnl<74$ at the 95\% confidence level. Constraints are expected to improve rapidly and significantly, first with Planck data and later using large scale structure data, see the recent reviews 
\cite{Komatsu:2010hc,Liguori:2010hx,Verde:2010wp}.
The Bardeen potential is related to the primordial curvature
perturbation of $\zeta$ on large scales and in the matter dominated era by
$\Phi=(3/5)\zeta$.

The curvature perturbation at horizon exit is  determined by the classical 
perturbations of the scalar fields, $\delta \phi_i({\bf x})$.
The subsequent evolution of $\zeta$ can be conveniently described by 
the $\delta N$ formalism~\cite{starob85,ss1,Sasaki:1998ug,lms,Lyth:2005fi}.
The curvature perturbation is given by up to quadratic terms~\cite{Lyth:2005fi}
\dis{\label{deltaN} 
\zeta=\delta N= \sum_I
N_{,I}\delta\vp_{I*}+\frac12\sum_{IJ}N_{,IJ}\delta\vp_{I*}\delta\vp_{J*}+\cdots\,. }
where $N({\bf x},t)$ is the e-folding number evaluated in an unperturbed
Universe,
from the epoch of horizon exit to later epoch of uniform energy density
hypersurface (for an extension to include gradient terms see~\cite{Takamizu:2010xy}).
The power spectrum $\calP_{\zeta}$ and the bispectrum $B_\zeta$ are defined by 
\begin{eqnarray}\label{powerspectrumdefn} \langle\zeta_{\bkone}\zeta_{\bktwo}\rangle &\equiv&
\picube\,
\sdelta{\bkone+\bktwo}\frac{2\pi^2}{k_1^3}\calP_{\zeta}(k_1) \, , \\
\langle\zeta_{{\mathbf k_1}}\,\zeta_{{\mathbf k_2}}\,
\zeta_{{\mathbf k_3}}\rangle &\equiv& \picube\, \sdelta{{\mathbf
k_1}+{\mathbf k_2}+{\mathbf k_3}} B_\zeta( k_1,k_2,k_3) \,. \end{eqnarray}
From this we can define the observable quantities, the spectral index, the
tensor-to-scalar ratio and the non-linearity parameter:
\bea n_{\zeta}-1&\equiv& \frac{\partial \log\calP_{\zeta}}{\partial\log k}, \label{tilt}\\
r&=&\frac{\calP_T}{\calP_{\zeta}}=\frac{8\calP_*}{\Mp^2\calP_{\zeta}}, \\
\fnl&=&\frac56\frac{k_1^3k_2^3k_3^3}{k_1^3+k_2^3+k_3^3}
\frac{B_{\zeta}(k_1,k_2,k_3)}{4\pi^4\calP_{\zeta}^2}, \label{fnldefn} \eea
where $\calP_T=8\calP_*/\Mp^2=8H_*^2/(4\pi^2\Mp^2)$ is the power spectrum of
the tensor metric fluctuations.
It is well known that single-field inflation does not lead to a detectably large 
non-Gaussianity, in fact $\fnl$ is suppressed by slow-roll
parameters~\cite{Maldacena:2002vr}. Observably large non-Gaussianity can be
obtained by breaking the slow-roll conditions during
inflation~\cite{Chen:2006xjb}, using extended kinetic terms~\cite{DBI}, see also the reviews~\cite{Koyama:2010xj,Chen:2010xk}, or
going beyond models of single-field
inflation~\cite{Bartolo:2003jx,Malik:2006pm,Lyth:2005fi,Zaldarriaga:2003my,Lyth:2005qk,preheating}

It is natural to consider multiple scalar field since they are ubiquitous in many beyond the standard model of
particle physics, such as supersymmetry and string theory. 
These scalar fields generate non-adiabatic perturbations during
inflation and change the evolution of the curvature perturbation after
horizon exit. The residual isocurvature perturbation may be present
in the primordial density fluctuation and can be correlated with the curvature
perturbation or may be responsible for an observably large 
non-Gaussianity in the cosmic microwave background and large scale 
structure, for observational limits on isocurvature perturbations see~\cite{Komatsu:2001rj,Valiviita:2009bp,Hikage:2009rt}. In this review we will only consider models with adiabatic primordial perturbations, in which the isocurvature perturbation present during inflation is converted into an adiabatic perturbation. We also neglect the secondary non-Gaussianities generated at later times, for example see~\cite{Bartolo:2010qu,Pitrou:2010sn}.

There are popular multi-field models~\cite{Wands:2007bd} which may generate observably large non-Gaussianity.
These include the curvaton scenario, modulated (p)reheating and an inhomogeneous
end of inflation, see Sec.~\ref{sec:summary}. In these scenarios, large non-Gaussianity is generated either by the means of ending inflation, or after inflation.
It was shown recently that it also possible to
generate large non-Gaussianity during the evolution of slow-roll multi-field inflation, see Sec.~\ref{slow-roll inflation}.

All of these models generate the large non-Gaussianity after horizon exit, such as after reheating,
at the end of inflation, at the phase transition or during inflation after horizon exit
and  involves the perturbation of the non-adiabatic mode. Therefore the
non-Gaussianity of these models is of the local type which is distinguishable from
other shapes of non-Gaussianity (for a list of possibilities see e.g.~\cite{Komatsu:2009kd,Fergusson:2008ra}),
in which the non-Gaussianity is generated intrinsically from the quantum fluctuations, or during horizon exit.

In Sec.~\ref{sec:summary} we summarise the aforementioned three models, which are popular methods of generating a large
non-Gaussianity. In Sec.~\ref{slow-roll inflation}, we review the possibility of generating a large non-Gaussianity 
from multi-field slow-roll inflation and in Sec.~\ref{sec:hybrid} we consider hybrid inflation with two inflaton fields as an application. Then in Sec.~\ref{sec:nonSR} we discuss multi-field models of inflation without assuming the slow-roll conditions and present an exact solution. Non-Gaussian observables beyond $\fnl$, such as its scale dependence and the trispectrum are introduced in Sec.~\ref{sec:beyondfnl}. Finally we conclude in Sec.~\ref{sec:conclusions}.

\section{Summary of popular models generating a large local non-Gaussianity}
\label{sec:summary}

Multi-field inflationary models~\cite{Wands:2007bd} can lead to an observable
signature of non-Gaussianity. These include models in which the large non-Gaussianity is generated either by the means of ending inflation, or after inflation. We review the three most popular of these models in this section, for a discussion on how these models are related see~\cite{Alabidi:2010ba}. In the next section we review the possibility of generating large non-Gaussianity 
during multi-field slow-roll inflation. Later in Sec.~\ref{sec:nonSR} we discuss multi-field models of inflation without assuming the slow-roll conditions. This does not exhaust all of the possible ways of generating a large local non-Gaussianity from inflation, see also the reviews~\cite{Mazumdar:2010sa,Bernardeau:2010jp,Wands:2010af,Tanaka:2010km} and the papers~\cite{Barnaby:2008fk,Langlois:2009jp}. Non-Guassianity of the local form can also be generated in the ekpyrotic scenario~\cite{Lehners:2009qu,Lehners:2010fy}, although in the simplest case $f_{NL}$ is large and negative~\cite{Koyama:2007if}, which is observationally ruled out.

\subsection{Curvaton scenario}

In this scenario there is a light 
weakly-interacting inhomogeneous scalar field, the curvaton $\chi$, in addition to the inflaton field, $\phi$.
The additional scalar field is completely subdominant during inflation, 
however it can dominate the energy density of the Universe later since
the time-averaged equation of state becomes that of the pressureless matter
(P=0), whose energy density decreases more slowly than that of 
the radiation which the inflaton field has decayed into. 
After decay the inhomogeneity of the scalar field leads to the density 
perturbation of radiation~\cite{Mollerach:1989hu,Linde:1996gt}. 
This is the basis of the curvaton 
scenario~\cite{Enqvist:2001zp,Lyth:2001nq,Moroi:2001ct}
which liberates the inflaton field from being required to generate the observed spectrum of perturbations (although there is instead an upper bound on their allowed magnitude).

The significant non-Gaussianity of the local type can be generated in the
curvaton model~\cite{Lyth:2002my}.
A study of non-Gaussianity in the curvaton scenario using second-order 
cosmological perturbation theory was done by Bartolo 
\etal~\cite{Bartolo:2003jx} with a sudden decay approximation and
also by Malik \etal~\cite{Malik:2006pm} using a fully numeric approach, which 
accounted for both the sudden decay and non-sudden decay approximations.
Lyth and Rodriguez~\cite{Lyth:2005fi} used the $\delta N$ formalism including
the quadratic terms (second order terms in \eq{deltaN})
to calculate the non-Gaussianity.
The non-linear generalisation of the curvature perturbation and a
numerical study was done to give a full probability
distribution function in~\cite{Sasaki:2006kq}.
This simple curvaton model was extended into the mixed curvaton-inflaton 
scenario in which fluctuations of both
the inflaton and a curvaton contribute to the primordial
density perturbation studying the isocurvature perturbation
in~\cite{Langlois:2008vk}. The multiple curvaton fields which contribute
to the primordial density perturbation was studied in~\cite{Choi:2007fya,Assadullahi:2007uw,Huang:2008rj}.
Observational constraints on curvaton models with isocurvature perturbation
and non-Gaussianity can be found in~\cite{Gordon:2002gv,:2008ei,Nakayama:2009ce}.

The generation of large non-Gaussianity in the curvaton scenario
can be easily understood using the result of the non-linear $\delta N$
formalism~\cite{Sasaki:2006kq,Langlois:2008vk}.
Assuming the sudden decay of the curvaton field on the $H=\Gamma$, i.e.,
when the local Hubble rate equals the decay rate for the
curvaton,
uniform energy-density hypersurface leads to a non-linear relation between the
local curvaton density and the radiation density before and after the decay,
\dis{
\Omega_re^{4(\zeta_{inf}-\zeta)}+ \Omega_{\chi}e^{3(\zeta_\chi-\zeta)}=1,
\label{omega_curvaton}
}
where $\Omega_r+ \Omega_{\chi}=1$. Here $\zeta$ is the primordial
curvature perturbation which remains constant
on large scales in the radiation-dominated era after the
curvaton decays and $\zeta_\chi$ is the curvaton perturbation.
Expanding this equation order by order yields up to second order~\cite{Sasaki:2006kq}
\dis{
\zeta= (1-r_\chi)\zeta_{inf}+ r_\chi\zeta_\chi +
\frac{r_\chi(1-r_\chi)(3+r_\chi)}{2}(\zeta_\chi-\zeta_{inf})^2+ {\rm higher-order\;terms},\label{zeta_curvaton}
}
where $r_\chi\equiv 3\Omega_\chi/(4-\Omega_\chi)$ at the time of the curvaton decay.
The curvature perturbation of curvaton field~\cite{Sasaki:2006kq}
\dis{
\zeta_\chi
= \frac23 \frac{\delta_1\chi}{\chi_0}  -\frac13\bfrac{\delta_1\chi}{\chi_0}^2
+\frac29\bfrac{\delta_1\chi}{\chi_0}^3
}
is almost constant before the curvaton decays from the start of curvaton
oscillation. Note that the intrinsic non-Gaussianity in the curvaton field
during oscillation is $f_{NL,\chi}=-\frac54$, which comes from the
non-linear relation between $\delta_1\chi$ and $\zeta_\chi$.
After the curvaton decays into radiation, the curvature perturbation of 
radiation has the following non-Gaussianity parameters:~\cite{Bartolo:2003jx,Lyth:2005fi,Sasaki:2006kq} 
\dis{
\fnl&=
\frac{\tr^2}{(1+\tr)^2}\left[\frac{5}{4r_\chi}-\frac{5}{3}-\frac{5r_\chi}{6}\right], \label{fnl-curvaton}
}
where $r_\chi$ is evaluated when the curvaton decay
 based on the sudden decay approximation but gives good agreement with a full
 numerical study~\cite{Malik:2006pm,Sasaki:2006kq}, and $\tr$ is defined as
\dis{
\tr\equiv \frac{|r_\chi\zeta_\chi|^2}{|(1-r_\chi)\zeta_{inf}|^2}.
}
In the limit of the pure curvaton scenario, $\zeta_{inf}\rightarrow 0$, it
recovers the usual normalization for $\fnl$, i.e. $\tr^2/(1+\tr)^2 \rightarrow 1$.
Therefore the large $\fnl$ is obtained for small $r_\chi$.
This large $\fnl$, in spite of the order of unity non-Gaussianity in $\zeta_\chi$,
has the origin in the non-linear relation between
$\zeta_\chi$ (or $\delta_1\chi$) and $\zeta$ in \eq{zeta_curvaton}.

In fact \eq{omega_curvaton}, and thus \eq{zeta_curvaton}, 
can be used ``at any time'' before the curvaton decay with $r_\chi$ evaluated at
that time to find the total curvature perturbation at that moment, as well as at the epoch of curvaton decay. This enables us to 
understand the evolution of the curvature perturbation $\zeta$ on the
uniform energy density hypersurface before the curvaton decay.
Initially when the primordial radiation was generated, e.g.~from the inflaton decay,
$r_\chi$ is negligibly small
so we cannot ignore the perturbations of the inflaton field.
At this early time, $\zeta= \zeta_{inf}$ and $\fnl=0$ since $\tr\propto
r_\chi^2$ which makes $\fnl$ negligible in \eq{fnl-curvaton}.
 As time goes on $r_\chi$ increases 
while $\zeta_{inf}$ and $\zeta_\chi$ are almost constant since there is no
energy flow between two fluids satisfying adiabatic condition except around the
decay time, thus the curvature perturbation evolves {\it gradually}. 
If the curvaton survives long enough, $r_\chi$ reaches a value
with which the $r_\chi\zeta_\chi$ dominates the first term, $\zeta_{inf}$,
and the curvaton perturbation is then responsible for the primordial inhomogeneity of the
Universe.
The large non-Gaussianity is possible only when $r_\chi$ at the decay time is 
small, i.e.~the energy density of curvaton is subdominant compared to
that of radiation.
This is because when $r_\chi=1$, the curvaton energy density dominates over the radiation component before decay,
then $\zeta$ is effectively the curvature perturbation of a single field
(curvaton), $\zeta_\chi$,
and therefore non-Gaussianity $\fnl=f_{NL,\chi}=-\frac54$.
We will also see this phenomena in the multi-field slow-roll inflation 
later, in Sec.~\ref{slow-roll inflation}.

\subsection{Modulated (p)reheating}
\label{modulated reheating}

After inflation the energy density in the inflaton field must be transferred into radiation. In the simplest case of adiabatic perturbations this process does not affect the primordial curvature perturbation on scales which are observable today, because these scale were much larger than the horizon at the time of reheating. However in a spirit similar to the curvaton model, there may be a subdominant light scalar field present during inflation which modulates the efficiency of reheating. This makes the efficiency of reheating a spatially dependent process. The quasi-scale invariant perturbations in this field, which during inflation are an isocurvature perturbation, may be converted into the primordial curvature perturbation during this process. For a review of reheating after inflation see for
example~\cite{Bassett:2005xm,Frolov:2010sz}.

As a simple illustration of this, in the ``old" models of perturbative
reheating the decay rate of the inflaton is given by $\Gamma\sim\lambda^2 m$, where $\lambda$ is the strength of the coupling between the inflaton and reheating fields and $m$ is the mass of the inflaton. Provided that $\Gamma$ is much less than the Hubble rate at the end of inflation, so that reheating takes place slowly, the reheating temperature is given by
\bea \Tr\sim\sqrt{\Gamma\Mp} \sim\lambda\sqrt{m\Mp}\,.\eea
If the coupling strength is a function of the local value of a second scalar
field $\chi$ then this will give rise to a perturbation in the decay
  rate of the inflaton field
and thus in the reheating temperature which is responsible for the density
perturbation after reheating. Provided that the $\chi$ field is
perturbed, the curvature perturbation can be written as
\bea \zeta=\alpha\frac{\delta\Gamma}{\Gamma}
\,, \eea
where $\alpha$ is a parameter depending on the ratio of $\Gamma$ to the Hubble
expansion at the end of inflation, whose value increases with decreasing
$\Gamma/H$ and leads to $1/6$ in the limit $\Gamma\ll H_{{\rm inf}}$~\cite{Dvali:2003em,Zaldarriaga:2003my}. Hence perturbations of the light scalar field $\chi$ can be imprinted into the radiation temperature through the reheating process~\cite{Dvali:2003em,Kofman:2003nx}. In a similar way the mass of the decaying particle may be modulated~\cite{Dvali:2003ar}. The above equation can be extended to second order, which is required to calculate $\fnl$.  In the simple quadratic case that $\Gamma=\Gamma_0+\Gamma_1(\chi/\bar{\chi})^2=\Gamma_0+\Gamma_1(1+\delta\chi/\bar{\chi})^2$ the result is~\cite{Zaldarriaga:2003my,Suyama:2007bg,Ichikawa:2008ne,Bauer:2005cd}
\bea \fnl=\frac{5}{12}\frac{\Gamma_0+\Gamma_1}{\Gamma_1\alpha}\,.\eea
Hence we can see that the non-Gaussianity can easily be larger than unity in
this model by two sources. One is due to the small value of $\alpha$ which
occurs when the decay rate of the inflation is not much smaller than the Hubble constant at the end
of inflation. The other is when $\Gamma_1$ is small in which $\chi$ field
only controls part of the decay channels of inflaton field.
. 

It is now generally considered that a period of rapid and highly non-linear preheating preceded the perturbative reheating phase~\cite{Bassett:2005xm}. In this case one can instead consider modulated preheating, see e.g.~\cite{BasteroGil:2003tj,Bond:2009xx,Kohri:2009ac}. A particularly rapid form of preheating, known as instant preheating~\cite{felder} has attracted more attention in this area, as it is possible to find analytic estimates~\cite{Kolb:2004jm,Byrnes:2005th,Matsuda:2006ee,Battefeld:2007st,Byrnes:2008zz}. It is possible to generate the primordial curvature perturbation in these models by having either a weak~\cite{Kolb:2004jm} or a strong~\cite{Matsuda:2006ee} symmetry breaking of the two-field potential during inflation which affects the efficiency of instant preheating, or by having two inflaton fields which are coupled with different strengths to the preheat field~\cite{Byrnes:2005th}. However all of the above models of modulated instant preheating are only allowed, even at the linear level in perturbations, for a small corner of parameter space and even then only for a special choice of initial conditions~\cite{Byrnes:2008zz}.

\subsection{Inhomogeneous end of inflation}\label{inhomogeneous end}

An inhomogeneous phase transition can occur also at the end of inflation
between inflationary and non-inflationary expansion due to the condition of how inflation ends~\cite{Lyth:2005qk,Huang:2009vk}.
In the case of multi-component inflation there is a family of inflationary trajectories.
If the family of trajectories is not perpendicular to the surface on which inflation ends then there is a possibility that inflation ends on a slice which is not of uniform density and the generation of the curvature perturbation at the end of inflation is expected~\cite{Lyth:2005qk}.~\footnote{This was also discussed
in~\cite{Bernardeau:2002jy,Bernardeau:2002jf,Bernardeau:2007xi} using the
linear relation between two fields at the end of inflation and the
non-Gaussianity is transferred from that of the light field which is generated
from its non-linear self-coupling.} 
In single field inflation the energy density is determined solely by the
single field $\phi$ and the end of inflation is parameterised by the unique 
value $\phi_e$. 
If there is another field $\sigma$, then $\phi_e(\sigma)$ will depend on
position through the perturbation $\delta\sigma({\bf x})$. As a result,
the change of e-folding number from a spacetime slice of uniform energy
density just before the end of inflation to a spacetime slice of uniform
density
just after the end of inflation will have a perturbation
 $\delta N_e=\zeta_e$. This $\zeta_e$ is the
  curvature perturbation generated by the end of inflation and can dominate over the
contribution from inflation~\cite{Lyth:2005qk}.

Using the perturbation of  $\phi_e(\sigma)$,
\dis{
\delta\phi_e=\phi_e'\delta\sigma + \frac12\phi_e''(\delta\sigma)^2,
} 
the curvature perturbation generated at the end of inflation is~\cite{Lyth:2005qk} 
\dis{
\zeta_e&=N_e'\delta\phi_e+\frac12N_e''(\delta\phi_e)^2\\
&=N_e'\phi_e'\delta\sigma + \frac{1}{2}\left[2N_e''\phi_e^{'2} + n_e'\phi_e''  \right](\delta\sigma)^2.\label{zetae}
}

When $\zeta_e$ dominates the perturbation generated during the inflation,
the power spectrum is given by
\dis{
\calP_{e}=\frac{\phi_e^{'2}}{2\epsilon_e}\left(\frac{H_*}{2\pi} \right)^2.
}
The non-Gaussianity can also be generated at the end of inflation
since the perturbation at the end of inflation is not required to be Gaussian. 
Thus we expect that  the observable
large non-Gaussianity can be generated in this process. From \eq{zetae} the non-Gaussianity is~\cite{Lyth:2005qk} 
\dis{
\fnl=-\frac53\sqrt{\frac{\epsilon_e}{2}}\frac{\phi_e''}{(\phi_e')^2}.
}
This was applied to the simple example of extended hybrid
 inflation using sudden end approximation~\cite{Alabidi:2006wa, Salem:2005nd}.

Recently Sasaki \etal~calculated analytically the curvature perturbation and the non-Gaussianity
 in the multi-brid inflation models~\cite{Sasaki:2008uc,Naruko:2008sq} from
 horizon exit to the end of inflation considering the general couplings to the
 waterfall field.
In general we cannot just ignore the curvature perturbation generated
during inflation. Furthermore that is also affected by the end of inflation.
A deeper understanding of non-Gaussianity both from the evolution during
multi-field hybrid inflation and effects from the end of inflation
is considered by Byrnes \etal~\cite{Byrnes:2008wi,Byrnes:2008zy}
which is summarised in the next section.

\section{Non-Gaussianity during slow-roll inflation}
\label{slow-roll inflation}
It is well known that the single field slow-roll inflation with canonical kinetic terms 
generates a non-linear parameter of the order
of the spectral tilt of the spectrum and thus too small to be observed~\cite{Maldacena:2002vr}.
On the other hand, in multi-field inflationary models the 
non-adiabatic perturbation may change this conclusion.
Furthermore even within slow-roll inflation an observably large 
non-Gaussianity can be generated if certain conditions are satisfied.

Rigopoulos \etal~introduced a formalism~\cite{Rigopoulos:2004gr,Rigopoulos:2005xx} to deal with non-linearity based on a system of fully non-linear equations for long wavelengths and applied it to
a two-field inflation model.
In the case of two-field quadratic inflation with two massive fields, 
an analytic calculation is possible with slow-roll conditions and this shows that
the nonlinear parameter is small~\cite{Alabidi:2005qi,Vernizzi:2006ve}.
The numerical calculation confirms the analytical prediction that it is possible to generate a narrow spike of large non-Gaussianity
while the inflaton trajectory turns a corner, but the non-Gaussianity decays  quickly after 
the corner~\cite{Rigopoulos:2005us,Vernizzi:2006ve}. This spike is due to the temporary jump in 
the slow-roll parameters~\cite{Yokoyama:2007uu,Yokoyama:2007dw}.

Recently an analysis based on the general analytic formula was done 
by Byrnes \etal~\cite{Byrnes:2008wi,Byrnes:2008zy} and they showed 
that it is possible to generate observable large 
non-Gaussianity even during slow-roll multi-field inflation models and 
explicitly wrote the conditions as well as concrete examples\footnote{There is
  a claim~\cite{Cogollo:2008bi} that the sizeable value of $\fnl$ can arise from a loop 
corrections.}. In this section we summarise the general conditions for two-field slow-roll inflation to generate a large non-Gaussianity and briefly give a specific example. In the next section we consider the multi-field hybrid inflation model
which has been studied in depth and we include the effects from the end of inflation.

\subsection{General formulas}
In the case of a potential which is either sum or product separable, 
it is possible to compute the curvature perturbation 
using the $\delta N$-formalism and slow-roll conditions,
and an analytic formula for the non-linearity parameter
can be obtained. The general formula for the non-linearity parameter
$\fnl$  was calculated in the case of a separable potential by sum~\cite{Vernizzi:2006ve} and this was later  extended to the non-canonical kinetic terms 
as well as to the separable potential by product~\cite{Choi:2007su}.
This has been further generalized to the arbitrary number of fields~\cite{Battefeld:2006sz}
and to the trispectrum~\cite{Seery:2006js}.

For a product separable potential, $W(\vp,\chi)=U(\vp)V(\chi)$,
we can find the number of e-foldings analytically in the slow-roll 
limit~\cite{GarciaBellido:1995qq},
\dis{
N(\vp_*,\chi_*)=-\frac{1}{\Mp^2}\int_*^e\frac{U}{U_\vp}d\vp
=-\frac{1}{\Mp^2}\int_*^e\frac{V}{V_\chi}d\chi.
}
Then the  power spectrum and the non-linear parameters can be 
calculated straightforwardly using the derivatives of e-folding
number~\cite{GarciaBellido:1995qq,Vernizzi:2006ve,Choi:2007su}.
 In the case of two inflatons with canonical kinetic terms,
the power spectrum and spectral index are found to be ~\cite{Choi:2007su}
\dis{
{\cal P}_\zeta = \frac{W_*}{24\pi^2\Mp^4 }\left(\frac{u^2}{\epstar} 
+ \frac{v^2}{\ecstar}\right),\label{spectrum_p}
}
\dis{
n_\zeta-1=-2\epsilon^* -4\frac{u^2\left(1-\frac{\eta_{\vp\vp}^*}{2\epstar}\right)
+ v^2\left(1-\frac{\eta_{\chi\chi}^*}{2\ecstar}\right)}{u^2/\epstar + v^2/\ecstar},\label{index_p}
} 
\dis{
r=16\left(\frac{u^2}{\epstar} + \frac{v^2}{\ecstar}\right)^{-1},\label{r_p}
}
where the slow-roll parameters are
\dis{
\label{epsilon}\epsilon_\vp
=\frac{\Mp^2}{2}\left(\frac{U_\vp}{U}\right)^2=\epsilon \cos^2\theta,\qquad
\epsilon_\chi
=\frac{\Mp^2}{2}\left(\frac{V_\chi}{V}\right)^2=\epsilon \sin^2\theta,
}
where $\theta$ is the angle between the adiabatic perturbation and one of the
fields and
\dis{\label{eta}
\eta_{\vp\vp}
=\Mp^2\frac{V_{\vp\vp}}{V},\qquad
\eta_{\vp\chi}
=\Mp^2\frac{U_{\vp}V_{\chi}}{W},\qquad
\eta_{\chi\chi}
=\Mp^2\frac{V_{\chi\chi}}{V}.
}
Note that the superscript ``*" and ``e" denotes that the values are evaluated at
horizon exit and at end of inflation respectively.
In the above $u$ and $v$ are evaluated at the end of inflation as
\dis{
u\equiv\frac{\epsilon_\vp^e}{\epsilon^e}= \cos^2\theta^e, \quad v\equiv\frac{\epsilon_\chi^e}{\epsilon^e} = \sin^2\theta^e.
}

The non-linearity parameter $\fnl$ becomes~\cite{Choi:2007su} 
\begin{eqnarray}
  \fnl&=& \frac{5}{6}
\frac{2}{\left( \frac{u^2}{\epsilon_\vp^*}
+ \frac{v^2}{\epsilon_\chi^*} \right)^2}
\left[
\frac{u^3}{\epsilon^*_\vp}
\left(1
 - \frac{\eta^*_{\vp\vp}}{2 \epsilon^*_\vp}
\right)
+ \frac{v^3}{\epsilon^*_\chi}
\left(1
 - \frac{\eta^*_{\chi\chi}}{2 \epsilon^*_\chi }
\right)
- \left( \frac{u}{{\epsilon^*_\vp}}
- \frac{v}{{\epsilon^*_\chi}}
\right)^2 \calA_P
\right] 
,
\label{fNL4product}
\end{eqnarray}
where
\begin{eqnarray}
\calA_P&\equiv&-\frac{\epsilon^e_\vp \epsilon^e_\chi}{(\epsilon^e)^2}
\left[
\hat\eta^e- 4 \frac{\epsilon^e_\vp \epsilon^e_\chi}{\epsilon^e}
\right],\\
\hat\eta &\equiv&
\frac{\epsilon_\chi\eta_{\vp\vp}+\epsilon_\vp\eta_{\chi\chi}}{\epsilon}.
\end{eqnarray} 
Similar formulas are found in the case of a sum separable potential~\cite{Vernizzi:2006ve,Choi:2007su}

\subsection{Conditions for generating a large $\fnl$}
Using an analytic formula for the nonlinear parameter $\fnl$ in the case of a sum or product separable potential,
it is possible to generate significant non-Gaussianity even during slow-roll inflation with Gaussian
perturbations at Hubble exit and the general conditions for it can be written explicitly~\cite{Byrnes:2008wi}.

For the separable potential by product large nonlinear parameter $|\fnl| \gtrsim 1$ is possible
when~\footnote{There is a symmetric region of large non-Gaussianity under the exchange of the two fields.}
\bea \sis\lesssim\sin^4\theta^e \left(\frac{1}{\sqrt{\ses \calG_p}}-1\right)\, ,
\qquad
 {\cal G}_p= \frac{6}{5}\left| -\eta^*_{\chi\chi} + 2 \eta^e_{\chi\chi} \right|^{-1}.\label{gen:Bconds}
\eea
With this condition, the large $\fnl$ is given by
\bea
\fnl &\simeq&
\frac{5}{6}
\frac{\sin^6\theta^e}{\left(\sin^2\theta^*+\sin^4\theta^e\right)^2}
\left[
-\eta^*_{\chi\chi} + 2 \eta^e_{\chi\chi}
\right].
\label{gapprox}
\eea
For large $\fnl$ the substantial increase in $\sin\theta$ is necessary between horizon exit and
the end of inflation, which corresponds to the curve of the trajectory in the field space
 during the slow-roll inflation.  However the background trajectory must be
 almost entirely in the one-field direction. 
 
One specific toy model is the quadratic times exponential potential~\cite{Byrnes:2008wi}
where the potential is given by
 \dis{W(\vp,\chi)=\frac12 e^{-\lambda\vp^2/\Mp^2}
m^2\chi^2.
} 
With expansion of exponential, this potential can be understood as the quadratic chaotic inflation model
which has a coupling to another light scalar field.
For this example large $\fnl$ is generated when the $\vp$ at horizon exit is small
$\vp^* \sim 10^{-3}\Mp$  for $\lambda=0.04$ and $\chi^*=16\Mp$.
 
For a sum separable potential, we can find similar conditions for generating a large $\fnl$~\cite{Byrnes:2008wi}.

\section{Hybrid inflation with two inflaton fields}
\label{sec:hybrid}

We consider a model of two field hybrid inflation, whose potential is given by
\bea\label{Whybrid}
W(\vp,\chi)= 
W_0\exp\left(\frac12\etap\frac{\vp^2}{\Mp^2}\right)\exp\left(\frac12\etac\frac{\chi^2}{\Mp^2}\right)\simeq
W_0\left(1+\frac12\etap\frac{\vp^2}{\Mp^2}+\frac12\etac\frac{\chi^2}{\Mp^2}\right)
\,, \eea
which is vacuum dominated, i.e.~which satisfies $\left|\etap\vp^2\right|\ll\Mp^2$ and
$\left|\etac\chi^2\right|\ll\Mp^2$. It is in this regime that the two ways of writing the potential given above are equivalent (at leading order in slow roll), so we can treat the potential as being both sum and product separable. 
We assume that inflation ends abruptly by
a waterfall field which is heavy during inflation and hence doesn't affect the dynamics during
inflation. First we calculate observables during slow-roll inflation. We will consider
the full potential including the waterfall field in Sec.~\ref{sec:waterfall}, which incorporate the effects from the end of inflation considered in Sec.~\ref{inhomogeneous end}. We will see that this
can lead to a change in observables on the surface where the waterfall field is destabilised.

In the vacuum dominated regime the slow-roll solutions are 
\dis{\label{hybridsoln}
\vp(N)= \vp_*e^{-\etap N}, \qquad \chi(N)= \chi_*e^{-\etac N},
}
where `*' denotes the value at the horizon exit. Throughout this section whenever we write a
quantity without making it explicit at which time it should be evaluated, we mean the equation to
be valid at any time $N$ $e$--foldings after Hubble exit and while slow roll is valid. Generally we
will be interested in quantities at the end of inflation, in which case we take $N=60$.

The slow-roll parameters are
\dis{\label{hybridsr}
\epsilon_\vp=\frac12\etap^2 \frac{\vp^2}{\Mp^2}
,\qquad
\epsilon_\chi=\frac12\etac^2 \frac{\chi^2}{\Mp^2}
, \qquad \epsilon= \epsilon_\vp+ \epsilon_\chi.
}
We note that the dominant slow-roll parameters $\etap$ and $\etac$ are
constants during inflation in the vacuum dominated regime and that they are much
larger than the slow-roll parameters $\epsilon_\vp$ and 
$\epsilon_\chi$ throughout inflation.

From the previous section, large non-Gaussianity can be realised
in either of two regions
\dis{
\cos^2\theta \equiv\frac{\dot{\vp}^2}{\dot{\vp}^2+\dot{\chi}^2}\simeq
\frac{\epsilon_\vp}{\epsilon_\vp+\epsilon_\chi } \ll 1,\qquad\textrm{or}\qquad 
\sin^2\theta \equiv \frac{\dot{\chi}^2}{\dot{\vp}^2+\dot{\chi}^2}\simeq
\frac{\epsilon_\chi}{\epsilon_\vp+\epsilon_\chi } \ll 1. 
}
Since the two regions are symmetrical~\cite{Byrnes:2008wi} (before specifying the values of $\etap$
and $\etac$), in the rest of this section we will focus on the second
region. In this region
where $\epsilon_\vp\gg\epsilon_\chi$,
$|\fnl|>1$ is fulfilled by the condition, 
\bea \sis\lesssim \sin^4\theta
\left(\sqrt{\frac{5|\etac|}{6\sin^2\theta}}-1\right)\, ,
 \label{fnlB}
\eea
in other words,
\dis{
|\etac|^{-1}e^{-4(\etap-\etac)N}  \lesssim \sin^2\theta  \simeq
\frac{\epsilon_\chi}{\epsilon_\vp}\lesssim |\etac|.
}
This implies three inequalities on $\theta$:
\bea 
\sis<\frac13\left(\frac{5}{6}\right)^2\left(\frac{3}{4}\right)^4\left| \eta_{\chi\chi} \right|^2,
\quad \sin^2\theta<\frac{5}{6}\left| \eta_{\chi\chi} \right|, \qquad
\frac{\sin^2\theta}{\sis}>\frac{24}{5}\frac{1}{|\etac|}.\label{threeconditions}
\eea
Note that in this region $\sin\theta\simeq\etac\chi/(\etap\vp)$, from Eq.~(\ref{hybridsoln}) we
require $N(\etap-\etac)>1$ so
that $\sin^2\theta$ grows significantly during inflation.

\subsection{Simplified formula for the observables when $\fnl$ is large}

We can substantially simplify all of the above formula in the case where $\fnl$ is large. We define
the quantity 
\dis{\label{trdefinition}
\tr\equiv \left(\frac{\p N}{\p \chi_*}\right)^2/ \left(\frac{\p N}{\p
\vp_*}\right)^2=\frac{\ec}{\ep} e^{2(\etap-\etac)N}.
}
In the region we are considering where $\fnl$ is large,
this is approximately given by the initial and final angles of the background trajectory with
different exponents
\bea \tr\simeq \frac{\sin^4\theta}{\sis}. \eea

In the case of large non-Gaussianity it follows that
\dis{
\calP_\zeta\simeq\frac{W_*}{24\pi^2\Mp^4\epsilon_*}\left(1+\frac{\ec}{\ep}e^{2(\etap-\etac)N}
\right)= \frac{8}{r}\left(\frac{H_*}{2\pi}\right)^2,\label{Pzeta_our}
}
\dis{
n_\zeta -1 \simeq 2\frac{\etap +\tr \etac}{1+\tr},
}
\dis{
r\simeq \frac{16\epsilon^*}{1+\tr},
}
\dis{\label{fNL_our}
\fnl\simeq
 \frac56 \frac{\sin^6\theta^e}{(\sis+\sin^4\theta^e)^{2}}\etac 
=\frac56\frac{\tr }{(1+\tr)^2}
\etac e^{2(\etap-\etac)N}.
}

The first condition  in Eq.~(\ref{threeconditions})  implies that
\bea \chis\ll\vps.  \eea
We therefore require a very small value of $\chis$ in order to have a 
large non-Gaussianity. While this may be a fine tuning, we comment that
requiring a large ratio of the initial field values is common in models where the isocurvature perturbation generates a significant part of the primordial adiabatic perturbation, see the discussion in the conclusion of~\cite{Byrnes:2008zz}. If the inflaton perturbations are neglected a priori, such as in the pure (i.e.~not mixed) curvaton scenario, this remains implicitly true. For an explicit discussion of this, in the simple case that both the inflaton and curvaton fields have a quadratic potential see \cite{Bartolo:2002vf}.

The sign of $\fnl$ is determined by the sign of $\etac$. 
The amplitude of $\fnl$ depends exponentially on the difference of the 
slow-roll parameters, $\etap-\etac$, which we
require to be positive to be in the branch of large non-Gaussianity 
where $\sin^2\theta\ll1$, while the spectral index depends on a weighted sum of the slow-roll 
parameters, so it is
possible to have a large non-Gaussianity and a scale invariant spectrum. However it is not
possible to have a large and positive $\fnl$ and a red spectrum of perturbations. We will see
in Sec.~\ref{subsecB} that by including the effect of the waterfall field this conclusion may
change, depending on the values of the coupling constants between the two inflaton fields and the
waterfall field.

In Table~\ref{table_hybrid}, we give some explicit examples of values of $\etap,\,\etac,\,\vps$ and
$\chis$ which
lead to a large non-Gaussianity. Using Eq.~(\ref{tilt}) we also calculate the
spectral index. The first example in the Table~\ref{table_hybrid} shows that it is possible to have
$|\fnl|\simeq100$ and a scale invariant spectrum. 
We also see that it is possible to generate a large non-Gaussianity during slow roll
with $\etap$ and $\etac$ both positive or both negative, or when one is positive and the other negative corresponding to a saddle point. The results for this model were verified using a novel calculational method of momentum transport equations in~\cite{Mulryne:2009kh}.
%
\begin{table}
\begin{tabular}{|cc|cc|cc|c|c|c|c|c|c|}
\hline
$\etap$ & $\etac$ & $\vps$ & $\chis$ & $\tr$ & $\fnl$ & $\tau_{NL}$ & $g_{NL}$ & $n_{\zeta}-1$ &   
r \\
\hline0.04 & -0.04   & 1 & 6.8$\times10^{-5}$ & 1 & -123  & 4.4$\times10^{4}$ & -33 & 0 & 0.006 \\ 
0.04 & -0.04 & 1 & $1.5\times10^{-4}$ & 5 & -68 & $8\times10^{3}$ & -24 & -0.05 & 0.002 \\
0.08 & 0.01    & 1 & 0.0018     & 1 & 9.27  & 247   & 0.77  & 0.09 & 0.026  \\ 
0.02 & -0.04   & 1 & 0.00037    & 1 & -11.1 & 357   &-2.6 &-0.02 & 0.002 \\
-0.01 & -0.09 &  1 & $3\times10^{-6}$ & 0.16 & -132 & $1.8\times10^{5}$ & -44 & -0.04 & 0.0007  \\
0.06 & -0.01 & 1 & $4.3\times10^{-4}$ & 0.1 & -3 & 148& -0.2& 0.11 & 0.026    \\
0.01 & -0.06& 1& $7.5\times10^{-6}$& 0.04 & -8 & 2.5$\times 10^{3}$ & -2 & 0.01 & 0.0008  \\
\hline
\end{tabular}
\caption{Table showing some initial conditions for the hybrid inflation model that lead to large
levels of non-Gaussianity. The table shows the parameter $\tr$, the bispectrum and the trispectrum non-linearity parameters, the spectral index and tensor-to-scalar ratio. They are evaluated when the number of e-foldings from the end of inflation 
is $N_k=60$. The trispectrum parameters $\tau_{NL}$ and $g_{NL}$ are defined and calculated in Sec.~\ref{sec:trispectrum}.
}
\label{table_hybrid}
\end{table}

\subsection{Effect of the waterfall field and further evolution after inflation}\label{sec:waterfall}

In this section we include the effects of the waterfall field $\rho$ which is required to end hybrid
inflation. Inflation ends when the waterfall field is destabilised, i.e.~when its effective mass
becomes negative. During inflation the waterfall field is heavy and it is trapped with a vacuum
expectation value of
zero, so we can neglect it during inflation.
The end of inflation occurs when the effective mass of the waterfall field is zero, which occurs on
a hypersurface defined in general by~\cite{Sasaki:2008uc,Naruko:2008sq},
\dis{
\sigma^2=G(\vp,\chi)\equiv g_1^2 \vp^2+g_2^2\chi^2,\label{endinflation}
}
which is realised by the potential $W(\vp,\chi)$, defined by Eq.~(\ref{Whybrid}), where $W_0$ is given
by
\dis{
W_0=\frac12G(\vp,\chi)\rho^2 + \frac{\lambda}{4}\left(\rho^2-\frac{\sigma^2}{\lambda} \right)^2.
}
Here $g_1$ ($g_2$) is the coupling between the $\vp$ ($\chi$) field and $\rho$
is the waterfall field with self coupling $\lambda$.
In general the hypersurface defined by this end condition is not a surface of uniform energy
density. 

As discussed earlier, this is an example of a model with an inhomogeneous end of inflation, 
i.e.~where inflation ends at
slightly different times in different places. 
It has also been shown for the hybrid potential we are considering 
that this can be used to generate a large amount of non-Gaussianity, 
for certain parameters values and fine tuning of the parameters
~\cite{Naruko:2008sq,Alabidi:2006wa}. However these papers 
concern the large non-Gaussianity generated
 at the end of inflation rather than during slow-roll inflation, by having a very
large ratio of couplings $g_1/g_2\ll1$. Here we consider the case where $g_1$ and $g_2$ have the
same order of magnitude with $g_1^2/g_2^2=\etap/\etac$ in sec.~\ref{subsecA}
 and  with $g_1^2= g_2^2$ in sec.~\ref{subsecB}.

\subsubsection{$g_1^2/g_2^2=\etap/\etac$}
\label{subsecA}

In this case we have chosen the coupling constants (which can satisfy $g_1^2<0$ and/or $g_2^2<0$)
such that the surface where the waterfall field is destabilised corresponds to a surface of uniform
energy density. This is because the end condition can be rewritten as 
$\sigma^2=2g_2^2/\etac(\frac12\etap\vp^2 + \frac12\etac\chi^2 ) \propto W$. In this case the value of all observable quantities such as the power
spectrum and non-Gaussianity are the same as those we calculated previously which were valid at the
final hypersurface of uniform energy density during inflation.

\subsubsection{$g_1^2= g_2^2$}
\label{subsecB}

In this case, the end of inflation given by the
condition in Eq.~(\ref{endinflation}) does not occur on a uniform 
energy density hypersurface~\cite{Naruko:2008sq}. We will show how the
non-Gaussianity is modified by the condition at the end of inflation in this
example. In general we expect there to be some modification to non-Gaussianity from the end of
inflation, except in the special case we considered in~\ref{subsecA}. We have checked in~\cite{Byrnes:2008zy} that the correction from the extra expansion which occurs from the surface on which inflation ends up to a surface of uniform energy density is small and therefore one can still use the $\delta N$ formalism in this case.

In this case the power spectrum and $\fnl$ in the limit of large non-Gaussianity are~\cite{Naruko:2008sq}
\dis{
\calP_\zeta\simeq\frac{W_*}{24\pi^2\Mp^4\epsilon_*}\left(1+\frac{\etap^2}{\etac^2}\frac{\ec}{\ep}e^{
2(\etap-\etac)N} \right)= \frac{8}{r}\left(\frac{H_*}{2\pi}\right)^2,\label{Pzeta_M}
}
\dis{
n_\zeta -1 \simeq 2\frac{\etap +\frac{\etap^2}{\etac^2}\tr \etac}{1+\frac{\etap^2}{\etac^2}\tr},
}
\dis{
r\simeq 16\epsilon^* \left(1+\frac{\etap^2}{\etac^2}\tr\right)^{-1},
}
\dis{\label{fNL_M}
\fnl\simeq
\frac56\frac{\frac{\etap^2}{\etac^2}\tr
}{\left(1+\frac{\etap^2}{\etac^2}\tr\right)^2} \etap
e^{2(\etap-\etac)N}.
}
We can see that the observables at the end of inflation are changed by the additional ratio
of $\etap^2/\etac^2$ before $\tr$. 
Furthermore, $\fnl$ has an additional factor of $\etap/\etac$.

In Table~\ref{table_hybridB} we show the values of $\fnl$, $n_{\zeta}-1$ and
$r$ with end condition $g_1^2=g_2^2$ for the same
parameter values that we used in Table~\ref{table_hybrid} where the inflation
ends on the uniform energy density hypersurface. 
The first two examples show that if
$\etap=-\etac$ then the observables are unchanged for two different end
conditions except that the sign of $\fnl$ is switched. The
second example in the table shows that in this case it is possible to have a red spectral index and
a positive value of $\fnl$. For many values of the initial parameters when $|\etap/\etac|\neq1$ the
magnitude of $\fnl$ decreases compared to Table~\ref{table_hybrid}, but the reverse can also happen, an example of this where $|\fnl|$ grows by more than an order of magnitude is shown in the final row of Tables~\ref{table_hybrid} and \ref{table_hybridB}.

\begin{table}
\begin{tabular}{|cc|cc|c|c|c|c|}
\hline
$\etap$ & $\etac$ & $\vps$ & $\chis$ & $\tr$ & $\fnl$  & $n_{\zeta}-1$ & r \\
\hline

0.04 & -0.04   & 1 & 6.8$\times10^{-5}$  & 1 & 123   & 0 & 0.006 \\ 
0.04 & -0.04 & 1 & $1.5\times10^{-4}$ & 5 & 68  & -0.05 & 0.002 \\
0.08 & 0.01 & 1 & 0.0018 & 1 & 4.59 &0.02 & 0.0008  \\ 
0.02 & -0.04 & 1 & 0.00037 & 1 & 3.5 &0.02 & 0.026 \\
-0.01 & -0.09 &  1 & $3\times10^{-6}$ & 0.16 & -0.2 & -0.02 & 0.0008  \\
0.06 & -0.01 & 1 & $4.3\times10^{-4}$ & 0.1 & 38 &0.01 & 0.006    \\

\hline
\end{tabular}
\caption{Same as Table~\ref{table_hybrid} but with different end condition, $g_1^2=g_2^2$ as 
used in sec.~\ref{subsecB}. }
\label{table_hybridB}
\end{table}

\subsubsection{Further evolution after inflation}

So far we have assumed a quick transition
to the radiation epoch at the end of inflation, thereby neglecting the dynamics of the
waterfall field. However if we consider the role of the waterfall field, then
after the waterfall field is destabilised there may be a further evolution 
of the primordial curvature perturbation, 
which will lead to a change of the observable parameters. This applies
to any model with an inhomogeneous end of inflation since there are
 isocurvature perturbations still present after the waterfall field is 
destabilised and inflation has ended. Further evolution
will depend on the details of reheating in a model dependent way.
 To the best of our knowledge this 
issue has not been considered in depth in any paper. If we assume an instantaneous transition to radiation domination 
(so a completely efficient and immediate decay of the waterfall and 
inflaton fields) then there will be no further change to the
observables as we have argued in the previous section. 
However this is clearly an idealised case. 

In the special case where the waterfall field is also light during inflation Barnaby and Cline
\cite{cline} have shown there is the possibility of generating a large non-Gaussianity during
preheating for certain parameter values. This is possible even if there is only one inflaton field
and the waterfall field present. However in this case inflation does not end abruptly when the
waterfall field is destabilised so this is not the scenario we have considered in this paper. A recent claim that even a heavy waterfall field may lead to a scale invariant local non-Gaussianity has been withdrawn~\cite{Mulryne:2009ci}.


\section{Multiple-field inflation without slow roll}\label{sec:nonSR}
  
In this section we demonstrate a new method 
 that provides, for certain classes of models, 
 analytical 
 expressions
for $f_{NL}$  valid in regimes
{\it beyond} a slow--roll approximation~\cite{Byrnes:2009qy}.
This method is based on the first--order Hamilton--Jacobi formalism developed by Salopek and Bond~\cite{SB},
which allows us to express inflationary observables in the
multi-field case, without having to  focus on a  slow--roll
regime (see~\cite{kinney} for
a similar
application of this formalism to the single-field case). 
    
 The main idea on which our arguments 
are based is     to focus on  inflationary multi-field trajectories
in  which the {\it Hubble rate}, and not the potential,
is separable. 
  This approach allows us to analytically
   study cases where the 
  non-Gaussian parameter
  $f_{NL}$ becomes large,  in regimes in which the slow--roll
parameters are enhanced after Hubble exit, but before inflation ends.
In particular, 
we provide analytical 
equations that express the non-linearity parameter
$\fnl$, in terms of quantities that generalise the slow-roll
parameters during inflation.  Using this method we demonstrate in the next subsection an exact solution of multi-field inflation which can give rise to a large non-Gaussianity due to a strong break down in slow roll shortly before inflation ends.

The first-order Hamilton-Jacobi equations
of motion are
\bea
H^2 &=& \frac{1}{3}\,W(\phi,\chi)
+\frac{2}{3 }\,\left[
\left( \frac{\partial H}{\partial \phi} \right)^2+
\left( \frac{\partial H}{\partial \chi} \right)^2
 \right]\,, \label{hubeq1}\\
\dot{\phi} &=& -2\,\frac{\partial H}{\partial \phi} ,
\hskip1cm\hskip1cm 
\dot{\chi} \,=\, -2 \,\frac{\partial H}{\partial \chi}\,.
\label{scaleq1} 
\eea

We concentrate on solutions satisfying
the following Ansatz for the Hubble parameter $H(\phi, \chi)$:
 \be\label{anshub}
H(\phi,\,\chi)\,=\,H^{(1)}(\phi)+ H^{(2)}(\chi)\,,
\ee 
 that is, we demand that it 
can be split  as a sum~\footnote{Hence we follow an  analysis 
related to~\cite{Vernizzi:2006ve}. In an analogous
way, one 
could also consider a situation in which the Hubble 
parameter splits as a {\it product} of single-field pieces~\cite{Choi:2007su}. The formula's in this section were extended to an arbitrary number of fields with a sum separable Hubble parameter in~\cite{Battefeld:2009ym}.}
of   two
pieces, each one depending on a single field.
Notice that this  is different with  respect to the requirement
of separable potentials usually done in the literature and earlier in this article.

It is convenient to define the following quantities:
\bea
\delta^{\phi}&=&\left( \frac{H^{(1)}_{,\phi}}{H}\right)^2
\hskip0.6cm,\hskip0.6cm\delta^{\chi}\,=\,\left( \frac{H^{(2)}_{,\chi}}{H}\right)^2\,,
 \\
\gamma^{\phi}&=& \frac{H^{(1)}_{,\phi \phi}}{H}
\hskip1.2cm,\hskip1.2cm\gamma^{\chi}\,=\,
\frac{H^{(2)}_{,\chi \chi}}{H}\,,
\eea
and $\delta\,=\,\delta^{\phi}+\delta^{\chi}$.  
Although their definition resembles the corresponding
 one for  the usual slow roll parameters
$\epsilon$ and $\eta$, they do {\it not} coincide with them
 when taking  a slow
roll limit~\footnote{\label{footnotesr} It is nevertheless
simple to work out the relation with the slow roll parameters,
defined by Eqns.~(\ref{epsilon}) and (\ref{eta}). In a slow-roll
regime, one finds
$$\delta^\phi\,\simeq\,\frac{\epsilon^\phi}{2}\,,\hskip0.7cm
\gamma^{\phi}\,\simeq\,\frac{\eta^{\phi\phi}-\epsilon^\phi}{2} \,.$$}. 
We
assume that the quantities $\delta$
 and $\gamma$ are much smaller than unity at Hubble exit
 $t=t_*$, in order
that we can use the $\delta N$ formalism. During inflation, by definition,
we have to ensure that the quantity
  $\epsilon_H\,\equiv\,-\dot{H}/H^2=2\delta<1$. The quantities $\gamma$ can however  become
much larger than unity during inflation.

Using similar techniques to those applied to a separable potential, one can calculate the spectral index and $\fnl$, the details were given in~\cite{Byrnes:2009qy} and the results are 
\bea\label{nzeta}
n_{\zeta}-1\,&=&\,-  4\delta_*-\,4\,\frac{u_H\left( 
1-\frac{\gamma_*^\phi}{ \delta_*^\phi}
u_H \right)+v_H \left(
1-\frac{\gamma_*^\chi}{ \delta_*^\chi}
v_H \right)}{\frac{u_H^2}{\delta_*^\phi} + 
\frac{v_H^2}{\delta_*^\chi} 
}\,, \\
 \label{keyresult}
\frac{6}{5}\,\fnl\,&=&\, 2\frac{
\frac{u_H^2}{\delta_*^\phi}\,\left( 1
-\frac{\gamma^{\phi}_*}{ \delta_*^\phi}
\,u_H \right)+
\frac{v^2}{\delta_*^\chi}\,\left( 1
-\frac{\gamma^{\chi}_*}{ \delta_*^\chi}
 \,v_H\right)
+ 2\left( \frac{u_H}{\delta_*^\phi}
-  \frac{v_H}{\delta_*^\chi}
\right)^2\,{\cal A}_H
}{\left( \frac{u_H^2}{\delta_*^\phi}
+  \frac{v_H^2}{\delta_*^\chi}
\right)^2}\,,
\eea
where the new symbols are defined by
\be
u_H\equiv \frac{H_*^{(1)}+Z_e}{H_*}\,, \qquad
v_H\equiv \frac{H_*^{(2)}-Z_e}{H_*}\,, \qquad Z_e=\left(H^{(2)}_e\delta^{\phi}_e-H^{(1)}_e\delta^{\chi}_e\right)/\delta_e\,,
\ee
\be
{\cal A}_H\,=\,-\frac{H_e^2}{H_*^2}\,\frac{\delta_e^\phi \delta_e^\chi}{
\delta_e}\,\left( \frac12-\frac{\gamma^{ss}_e}{\delta_e}\right)\,,
\label{defcalA} \qquad  
\gamma^{ss}\,=\,\left(\delta^\chi \gamma^\phi+ 
\delta^\phi \gamma^\chi\right)/
\delta\,.
\ee

We reiterate that the formula for $\fnl$ is exact (apart from neglecting a slow-roll suppressed contribution due to the non-Gaussianity of the fields at Hubble exit) and is not
based on a slow-roll expansion. 
Although  $\delta_*$ and $\gamma_*$
are necessarily small, the quantity ${\cal A}$ is not suppressed by these 
parameters, and can assume large values  during inflation,
enhancing $\fnl$. We note in agreement with~\cite{Vernizzi:2006ve} that if one of the fields
has reached a minimum so $\dot{\phi}=0$ or $\dot{\chi}=0$ at the end of inflation then
$\mathcal{A}_H=0$ and the non-Gaussianity at the end of inflation will be very small. Therefore any model
of inflation with a separable potential or separable Hubble factor with a large non-Gaussianity
present at the
end of inflation must have both fields still evolving, and therefore the presence of isocurvature
modes. It would therefore also be interesting to study the evolution of the perturbations after
inflation, to see if this leads to an important change in the observables~\cite{Choi:2008et}.

\subsection{Exact solution}\label{sec:exactsolution}
Very few exact solutions in multiple field inflation are known~\cite{Sasaki:2007ay}. Here we present an exact solution which can give rise to a very large non-Gaussianity at the end of inflation.
A potential with a similar 
form 
may be motivated in string theory, in the context
of K\"ahler moduli inflation~\cite{kahlerinfl}.

The potential we consider is~\cite{Byrnes:2009qy}
\be
W(\phi,\chi)\,=\,U_0\,\left( 1-A_1 e^{-\alpha \phi}
+A_2  e^{-2 \alpha  \phi}
-B_1 e^{-\beta \chi}+
B_2 e^{-2\beta \chi}+
\frac{ A_1  B_1}{2}\,e^{-\alpha \phi-\beta \chi}\right)\,,
\ee
where the parameters
$\alpha$ and $\beta$
satisfy 
\bea\label{alphabetaconstraints}
\alpha^2 =\frac32-\frac{6 A_2}{A_1^2}\,, \qquad \beta^2 =\frac32-\frac{6 B_2}{B_1^2}\,.
\eea

The Hamilton-Jacobi equations have the following exact 
 solution  for $H$ and the fields:
\bea
H&=&H_0\,\left(1-\frac{A_1}{2}
 e^{-\alpha \phi}-\frac{B_1}{2} e^{-\beta \chi}\right)\,, \\
\phi&=&\frac{1}{\alpha}\,\ln{\left[e^{\alpha \phi_{*}}
-A_1 \,\alpha^2\,H_0\,t
\right]}\,, \qquad \chi=\frac{1}{\beta}\,\ln{\left[e^{\beta \chi_{*}}
-B_1 \,\beta^2\,H_0\,t
\right]}\,,
\eea
where we have defined $H_0=\sqrt{U_0/3}$, $t$ is cosmic time and we have set $t_{*}=0$. The overall factor to the potential $U_0$ can be freely chosen so that the amplitude of the scalar power spectrum matches the observed amplitude of perturbations in the CMB.
The scale factor results:
\be
a(t)\,=\,a_0\,\left(e^{\alpha\phi_{*}}-
  \alpha^2\, A_1\,H_0\,t
\right)^{\frac{1}{2 \alpha^2}}\,\left(e^{\beta\chi_{*}}-
  \beta^2\, B_1\,H_0\,t
\right)^{\frac{1}{2 \beta^2}}\,e^{H_0\,t}\,.
\ee
Notice that the solution becomes singular at late times, when the scale 
factor vanishes and the field values diverge. This singularity occurs well after 
inflation ends, and we will not need to discuss it in our analysis. From the relations
\bea
\gamma^{\phi}&=&-\alpha\,\sqrt{\delta^\phi}\hskip0.5cm,\hskip0.5cm
\gamma^{\chi}\,=\,-\beta\,\sqrt{\delta^\chi} \,,\label{relgam}
\eea
we notice that in this example $\gamma^{\phi}$ and $\gamma^{\chi}$ can become much larger than unity at the end of
inflation if $|\alpha|$ and/or $|\beta|$ are much greater than one. This corresponds to a break down in
slow roll, although $\epsilon_H$ remains smaller than unity during the inflationary era by definition.

As a simple, concrete example of parameter choice which gives rise to a large non-Gaussianity, we choose a regime in which
the parameters $|\alpha|$ and $|\beta|$ are both large,
let us say larger than some quantity $R\gg1$. We note from (\ref{alphabetaconstraints}) that we
are therefore required to have $A_2<0$ and $B_2<0$. In this case the potential does not have a
minimum, but inflation still ends through $\epsilon_H$ growing larger than unity, and we can trust
our results in this regime. The potential would need modifications which apply after inflation in
order for reheating to take place and these modifications may provide a minimum for the potential. We parameterise
the values of the quantities $\delta_e^\phi$ and $\delta_e^\chi$
at the end of inflation 
as 

\be\delta_e^\phi=\frac{1}{2 m}
\hskip0.4cm, \hskip0.4cm
\delta_e^\chi=\frac{1}{2 n}
\hskip0.8cm {\rm such \hskip 0.2cm that}
\hskip0.8cm
\frac{1}{m}+\frac{1}{n}=1\,.\label{fincondelta}\ee
where the final equality follows from $\epsilon_H=1$. We assume that the quantities $m$ and $n$
are not too big. Namely, they satisfy
the inequality $m,\,n\,\ll R^{2}$. This implies
 that $H_0\simeq H_* \simeq H_e$. 
We choose
 $|\gamma_*^\phi|=\frac{1}{40 p}$,  
  $|\gamma_*^\chi|=\frac{1}{40 q}$, such that we can write
 \be
 N_{tot}=60\,\simeq\,10\, \left(p+q \right)\,.
 \ee
This fixes the initial values for the fields
 $\phi_*$ and $\chi_*$,
 to the values
 \bea
 e^{\alpha \phi_*}&=& 20\, p\,\alpha^2 A_1 
 \label{inval1}\,,\\
 e^{\beta \chi_*}&=& 20\, q\,\beta^2 B_1\,.
 \label{inval2}
 \eea
 
The conditions (\ref{fincondelta}) imply
\bea H_0\,t_e &=& 20 p-\sqrt{\frac{m}{2\alpha^2}}\,,\\
H_0\,t_e &=& 20 q-\sqrt{\frac{n}{2\beta^2}}\,.
\eea
Since $m$ and $n$ are much smaller than $R^2$, we
expect $p\simeq q$.
We hence have the
simple relations
\bea \delta^{\phi}_{*}\simeq\frac{1}{4N_{tot}^2\alpha^2}, \qquad
\delta^{\chi}_{*}\simeq\frac{1}{4N_{tot}^2\beta^2}, \qquad
\gamma^{\phi}_{*}\simeq\gamma^{\chi}_{*}\simeq-\frac{1}{2N_{tot}}\,. \eea
From (\ref{nzeta}), we find
\bea n_{\zeta}-1&\simeq&-\frac{2}{N_{tot}}\simeq-0.04\,. \eea
So we have a red spectral index in
agreement with present day observations~\cite{Komatsu:2008hk} and in~\cite{Byrnes:2009qy} it was also shown that the tensor-to-scalar ratio for this model is negligible.
\begin{figure}
\begin{center}
\scalebox{1.2}{\includegraphics*{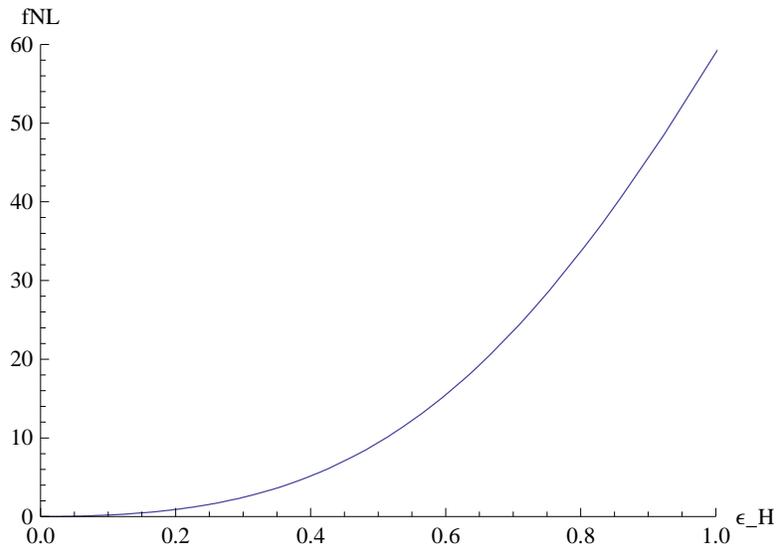}} 
\caption{Plot showing $f_{NL}$ as a  
 function of $\epsilon_H$ towards
 the end of
 inflation, for the values of the parameters given in the text. Inflation ends
 when $\epsilon_H=1$; for this example $f_{NL}\simeq59$ at that time.}\label{plot}
 \end{center}
\end{figure}

When evaluated at the end of inflation, we find
\bea\label{fNLconcrete}
\frac65 \fnl &\simeq&
 - 2\sqrt{2m n}\,\left[ \frac{\alpha^2 n-\beta^2 m}{\alpha^2 n^2
 +\beta^2 m^2}
 \right]^2
\,\left( \frac{\alpha}{\sqrt{n}}+
\frac{\beta}{\sqrt{m}}
 \right) \,.\label{fNLconcrete2} \eea
We stress that this formula provide only
the dominant 
contributions to $f_{NL}$. It is valid
 in the case
$|\alpha|,|\beta| > R\gg1$, and $m,n< R^2$.

\begin{figure}
\begin{center}
\scalebox{.5}{\includegraphics*{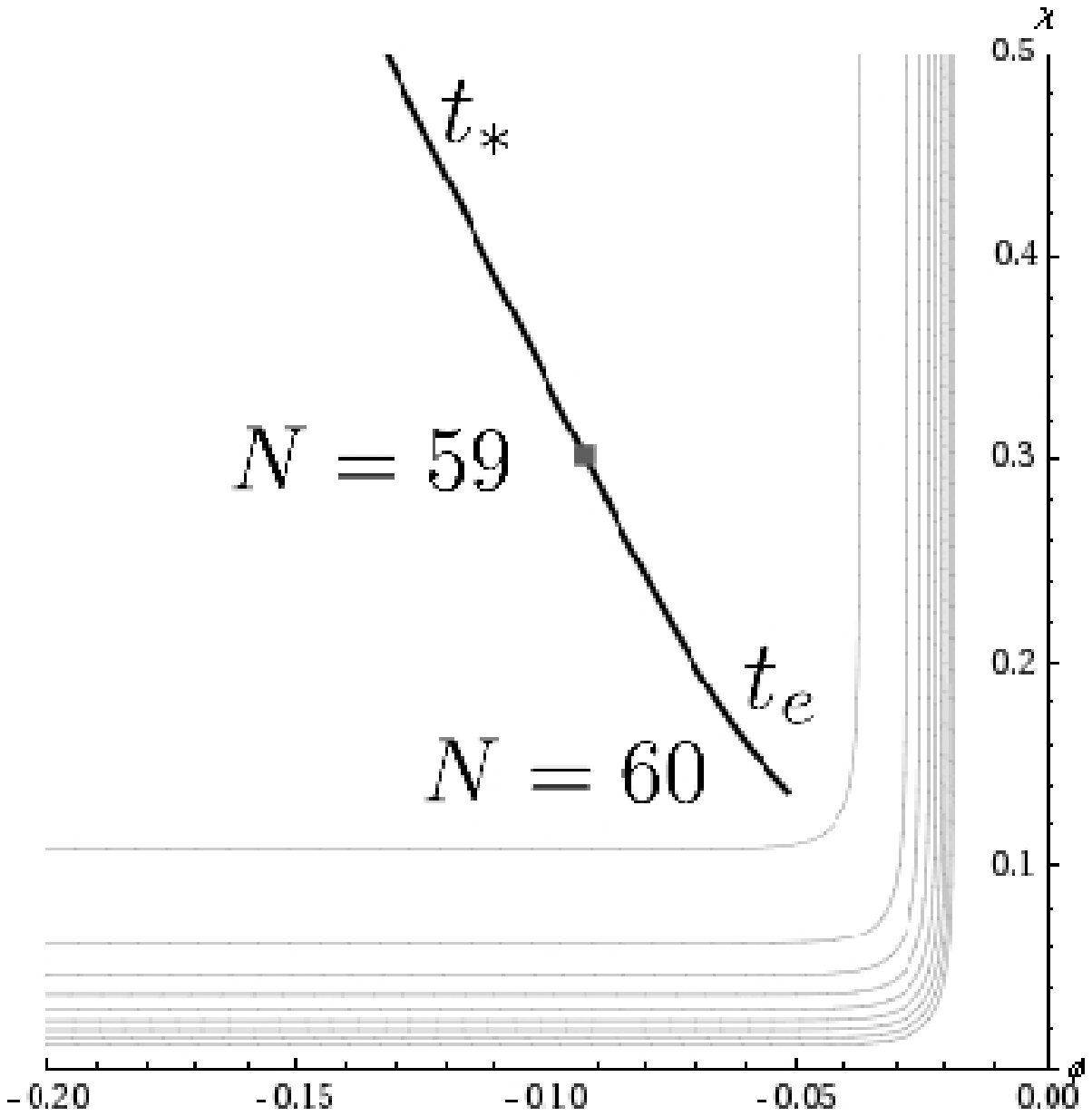}} 
$\qquad\qquad$
\scalebox{.7}{\includegraphics*{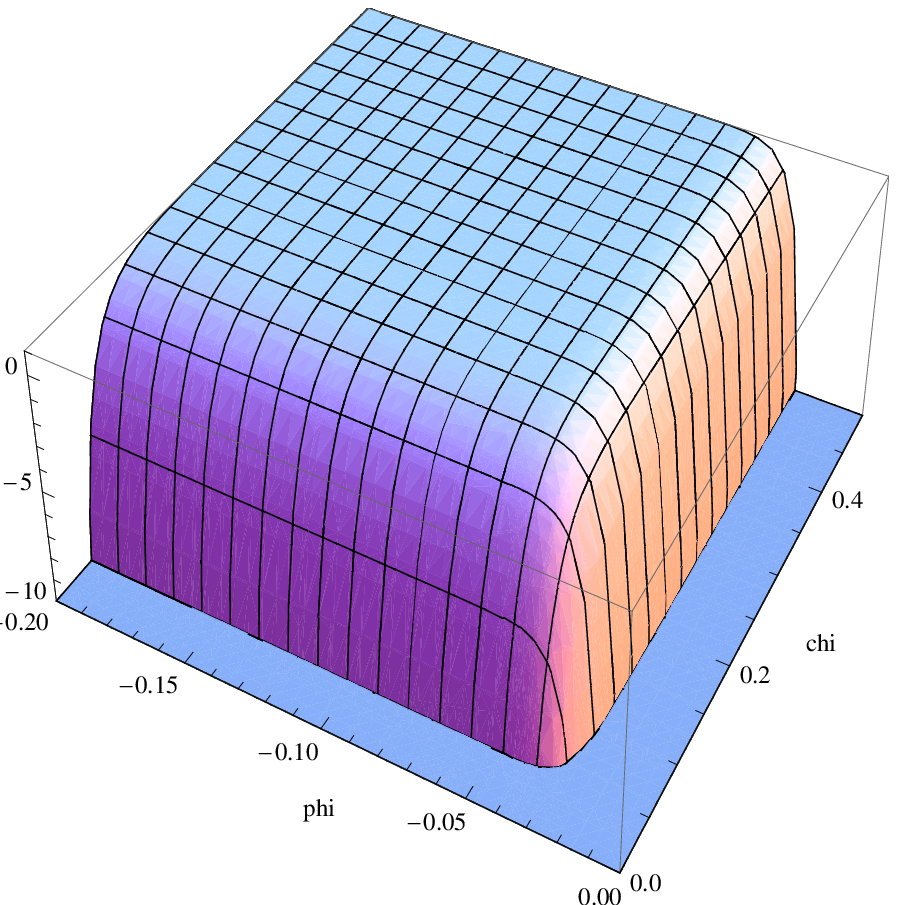}} 
\caption{Left plot shows the trajectory considered for the parameters given after
eq.~(\ref{fNLconcrete2}) superimposed on a contour plot of the potential. The square on the
trajectory indicates a point along the trajectory one $e$-folding before inflation ends as $\phi$
and $\chi$ roll towards zero. This shows that the fields roll much more quickly during the final 
stage of inflation, and the trajectory curves near the end. The right plot shows the potential for the same parameter values. Notice that
inflation ends on the plateau long before the potential becomes negative.}\label{Traj}
 \end{center}
\end{figure}

As an explicit case,  we take $\alpha=-100,\beta=20,m=6,n=6/5$ and the initial conditions to
satisfy (\ref{inval1}) and (\ref{inval2}) with $N_{tot}=60$. 
Then, independently of the values of
$A_1$ and $B_1$, we find $f_{NL}\simeq58$ from the simplified formula (\ref{fNLconcrete}), and
$f_{NL}\simeq53$ 
from the full
formula (\ref{keyresult}). See figures \ref{plot} and \ref{Traj}. Notice that the results for
$f_{NL}$ differ from
each other by quantities of order $1/R=1/\beta$ as expected by the approximations we have made.
At the end of inflation, the parameters are
$\gamma^{\phi}=29$, $\gamma^{\chi}=-13$ and this shows that the slow-roll approximation has been significantly broken, see footnote \ref{footnotesr}. From Fig.~\ref{Traj} one can see that the trajectory is straight for much of inflation but it turn near the end of inflation (during the last e-folding before inflation ends), and it is during this time that $\fnl$ grows larger than unity.

Notice that our requirements of final values for the quantities
$\delta_e^\phi$ and $\delta_e^\chi$ (both much bigger than
$1/R$) imposes fine-tuning constraints on the initial 
conditions, since at leading order in $1/R$ the values
for $\gamma_*^\phi$ and
$\gamma_*^\chi$ must coincide
 (see Eqs.~(\ref{inval1})-(\ref{inval2})
and recall that $p\simeq q$).

We stress that the enhancement of non-Gaussianity occurs towards the end of inflation. What happens
just {\it after} inflation is a model dependent issue; we cannot address this question within the
approximations used in this concrete model. For our choice of parameters we notice that $f_{NL}$ is
still increasing at the end of inflation, see figure \ref{plot}, but for other choices it may start
to decrease before inflation ends, see~\cite{Battefeld:2009ym}. We stress that in the regime where $|\alpha|,|\beta|\gg1$ the
slow-roll parameters will necessarily become much greater than unity by the end of inflation, which
may correspond to $|\gamma^{ss}|\gg1$: this is one reason by which $f_{NL}$ can become large. Our
formalism is at least in principle suitable to study the evolution of non-Gaussianity after
inflation, when the parameter $\epsilon_H$ becomes larger than unity (although we restate that the specific potential being considered here anyway needs to be modified after inflation ends).

\section{Higher order non-Gaussian observables}\label{sec:beyondfnl}

Non-Gaussianity is most popularly parameterised in terms of one non-linearity parameter, $\fnl$. There are several different $\fnl$'s used in the literature which parameterise different shape dependences of the bispectrum~\cite{Fergusson:2008ra}, the most popular being the local model discussed in this article and the so-called equilateral model which can arise from models of inflation with non-canonical kinetic terms (a popular example is DBI inflation~\cite{DBI}). But can we learn more than one number from an observation of non-Gaussianity? And if $\fnl^{\rm local}$ is detected how can we hope to distinguish between the models which can generate this? Fortunately observations of non-Gaussianity have the possibility of teaching us much more than the value of one parameter. Two examples of further non-Gaussian observables are a scale dependence of $\fnl$ and the trispectrum.

\subsection{Trispectrum}\label{sec:trispectrum}

The four point function of the primordial curvature perturbation is defined by
\begin{equation}\label{Tdefn}
\la \zeta_{{\mathbf k_1}}\,\zeta_{{\mathbf k_2}}\, \zeta_{{\mathbf k_3}} \zeta_{{\mathbf
k_4}} \ra_e \equiv T_\zeta({\mathbf k_1},{\mathbf k_2},{\mathbf k_3}, {\mathbf k_4})
\picube \sdelta{{\mathbf k_1}+{\mathbf k_2}+{\mathbf k_3} +{\mathbf k_4}}\,,
\end{equation}
which using the $\delta N$ formalism, and assuming that the fields have a Gaussian distribution at Hubble exit, is given by
\begin{align}\label{tauNLgNLdefn} T_\zeta (\bkone,\bktwo,{\mathbf k_3},{\mathbf k_4}) =&
\tau_{NL}\left[P_\zeta(k_{13})P_\zeta(k_3)P_\zeta(k_4)+(11\,\,\rm{perms})\right] \nonumber \\
&+\frac{54}{25}g_{NL}\left[P_\zeta(k_2)P_\zeta(k_3)P_\zeta(k_4)+(3\,\,\rm{perms})\right]\,,
\end{align}
where $k_{13}\equiv|{\bf k}_1+{\bf k}_3|$, and the trispectrum non-linearity parameters can be calculated using~\cite{Alabidi:2005qi,Seery:2006js,Byrnes:2006vq}
\bea
\label{tauNL}
\tau_{NL}&=&\frac{N_{AB}N^{AC}N^BN_C}{(N_DN^D)^3}\,, \\
\label{gNL}
g_{NL}&=&\frac{25}{54}\frac{N_{ABC}N^AN^BN^C}{(N_DN^D)^3}\,.
\eea
Hence we see that the trispectrum depends on two non-linearity parameters (as opposed to one, $\fnl$, for the bispectrum), and they may be observationally distinguishable since they are prefactors of terms with different shape dependences (\ref{tauNLgNLdefn}). The current observational bound on
the local type of the bispectrum from seven years of WMAP data is $-10<\fnl<74$ at the $2\sigma$ level
\cite{Komatsu:2010fb}. Recently there has been a first constraint on both $\tau_{NL}$ and $g_{NL}$, which came from using WMAP5 data, with the bounds $-7.4<g_{NL}/10^5<8.2$ and $-0.6<\tau_{NL}/10^4<3.3$ at $95\%$ confidence~\cite{Smidt:2010sv}. There have been two other observational constraints on the trispectrum through $g_{NL}$, setting $\tau_{NL}=0$. The bounds, which in both cases are roughly $|g_{NL}|\lesssim10^5-10^6$, come from large scale structure~\cite{Desjacques:2009jb},
and the CMB~\cite{Vielva:2009jz}. If
there is no detection of non-Gaussianity it is expected that with Planck data the bounds will be reduced to about
$|\fnl|\lesssim10$, $\tau_{NL}\lesssim10^3$ and $g_{NL}\lesssim10^5$ at the $2\sigma$ level and future observations may become even tighter~\cite{Kogo:2006kh,Smidt:2010ra}. Note that the expected future constraint on $g_{NL}$ is about two orders of magnitude weaker than that on $\tau_{NL}$~\cite{Smidt:2010ra}. We will see that it is also possible for some models and parameter ranges that the first observational signature will come through the trispectrum rather than the bispectrum.

\subsubsection{Trispectrum in two-field hybrid inflation}

In the regime where $|g_{NL}|$ and $\tau_{NL}$ are greater than unity they are given by
\bea  g_{NL}&=&\frac{10}{3}\frac{\tr\left(\etap-2\etac\right)-\etac}{1+\tr}\fnl, \\ \label{NLtrispectrum}
\tau_{NL}&=&\frac{\tr}{(1+\tr)^3}\etac^2e^{4N(\etap-\etac)} = \frac{1+\tr}{\tr}
\left(\frac{6}{5}\fnl\right)^2.  \eea
We see that $g_{NL}$ is subdominant to $\fnl$ and
hence won't provide a competitive observational signature. It follows from (\ref{NLtrispectrum}) that $\tau_{NL}> (6\fnl/5)^2$, so
$\tau_{NL}$ may be large and provide an extra observable parameter for this model. This
inequality between $\tau_{NL}$ and $\fnl$ is true in general~\cite{Suyama:2007bg}, and equality is
reached whenever a single field direction during inflation generates the 
primordial curvature perturbation. However it is usually assumed that $\tau_{NL}\sim\fnl^2$
since both arise from second derivatives in the $\delta N$ formalism. In fact for our model it is
possible to have a small $\fnl$ (and hence also a small $g_{NL}$) but a large and potentially
observable $\tau_{NL}$. For this we require that $\tr\ll1$, although in practice if we make it too small it
may no longer be possible to satisfy a constraint on a minimum possible field velocity consistent with a classical slow-roll trajectory, as discussed
in~\cite{Byrnes:2008zy} (see also~\cite{Creminelli:2008es} and a discussion which reaches a different conclusion is given in~\cite{Rodriguez:2008hy}). In the final example in Table~\ref{table_hybrid} we give an explicit
example of parameter values which give rise to an $\fnl$ which is probably too small to be detected
with Planck but with a very large trispectrum through $\tau_{NL}>10^3$ that should be detectable at
a high significance. 
For another example with $f_{NL},g_{NL}\lesssim O(1)$ but $\tau_{NL}\gg1$, see
~\cite{Ichikawa:2008ne}.
In contrast it has been shown in several papers
\cite{Enqvist:2005pg,Byrnes:2006vq,Huang:2008bg,Huang:2008zj,Enqvist:2008gk,Enqvist:2009ww} that in the curvaton
scenario where the curvaton has a non-quadratic potential it is possible to realise $|g_{NL}|\gg1$
while $\tau_{NL}=(6\fnl/5)^2$ is small with some tuning of parameters. This is also possible in the exact solution which we presented in the previous section and we consider this next.

\subsubsection{Trispectrum in the exact solution}

In this regime where $|\alpha|\gg|\beta|>R$ and $m,n>1/R^2$ it is also
possible to give compact expressions for the trispectrum (4-point function) non-linearity
parameters, in terms of
$f_{NL}^2$ as
\bea \tau_{NL}\simeq\left(\frac65 f_{NL}\right)^2, \qquad \frac{54}{25}g_{NL}\simeq-\frac32
\frac{n-m}{m}\left(\frac65 f_{NL}\right)^2. \eea
The complete result was given in~\cite{Battefeld:2009ym}. Hence both of the trispectrum non-linearity parameters are generally large whenever $\fnl$ is. We note that if $m\simeq1$ then from
(\ref{fincondelta}) $n\gg1$, so the trispectrum through $g_{NL}$ will give the dominant signal of
non-Gaussianity through a large, negative $g_{NL}$.

\subsection{Scale dependence of $\fnl$}

In its simplest form, the local form of $\fnl$ defined by Eq.~(\ref{Phifnl}) is a constant parameter, independent of both position and scale. However in realistic models $\fnl$ is likely to be mildly scale dependent. This may happen in two ways. In general single-field models (such as the curvaton scenario) where the scalar field which generates the primordial curvature perturbation has a non-quadratic potential the non-linearities this generates will give rise to scale dependence of $\fnl$. Alternatively even in models where all of the fields have a quadratic potential; if the primordial curvature perturbation has contributions from more than one field, and the fields don't all have the same mass, then $\fnl$ will again have a scale dependence. This is because the correlation between the first and second order terms of $\zeta$, which the bispectrum depends on, will become scale dependent. This is indeed what happens in the two-field hybrid model presented earlier. For detail
 s about a scale dependence of local $\fnl$ see Byrnes {\rm et al}~\cite{Byrnes:2009pe}.

We define a second observable parameter derived from the bispectrum
\bea n_{\fnl}=\frac{d\ln|\fnl|}{d\ln k}\,,\eea
which is analogous to the scale dependence of the power spectrum. There is a subtlety here, which is that in general $\fnl$ may depend on three independent parameters, $k_1,k_2$ and $k_3$. However it was shown in~\cite{Byrnes:2009pe} that provided one takes the derivative while keeping the ratio of the three $k$ vectors fixed then $n_{\fnl}$ is independent of the shape of the triangle described by the three $k$ vectors, which makes this a well defined quantity. Observational prospects for this quantity were considered in~\cite{Sefusatti:2009xu}, who showed that the Planck satellite is sensitive to $n_{\fnl}\simeq0.1$ assuming a fiducial value of $\fnl=50$. This observational sensitivity to $n_{\fnl}$ is about a factor of two larger than the current preferred value of the power spectrums spectral index and may therefore provide an interesting extra constraint on non-Gaussian models. The scale dependence of an equilateral form of $\fnl$ has also been considered from both a theoretical
  and observational perspective, see for example~\cite{Chen:2005fe,LoVerde:2007ri,Khoury:2008wj,Sefusatti:2009xu}.

In the example of two-field hybrid inflation $\fnl$ has a scale dependence both because of the exponential term in $\fnl$,
(\ref{fNL_our}), and
because $\tr$ will vary through the change of the initial value of $\sis$. We find 
\bea \frac{\partial \ln \tr}{\partial \ln k}=\frac{\partial \ln e^{2 N(\etap-\etac)}}{\partial \ln
k} =-2(\etap-\etac). \eea
Using this we find from (\ref{fNL_our}) that
\bea n_{\fnl} \equiv \frac{d\log \fnl}{d\log k}=
-4\frac{\etap-\etac}{1+\tr}. \eea 
In the case that we include the effect from the surface where the waterfall field is destabilised
and $g_1^2=g_2^2$ we find from (\ref{fNL_M}) that
\bea n_{\fnl}= -4\frac{\etap-\etac}{1+\left(\frac{\etap}{\etac}\right)^2\tr}. \eea
For both cases the spectral index of $\fnl$ satisfies 
\bea -4(\etap-\etac)<n_{\fnl}<0, \eea
 for any value of $\tr$ and hence $\fnl$
will be smaller on small scales.

Because we require a relatively large value of $\etap-\etac>1/N$ for our model to generate a large
non-Gaussianity it is quite possible for our model to generate a relatively significant scale
dependence of $\fnl$. However the amount also depends on $\tr$ and when this is large then the $\chi$ field is almost solely responsible for generating $\zeta$ at both first and second order and $n_{\fnl}$ is suppressed. We can also see in agreement with the statement at the beginning of this subsection that in this case if the mass of the two fields are equal then the two fields have the same scale dependence and $n_{\fnl}=0$.

We note that this is in contrast to the large non-Gaussianity from an inhomogeneous end of
inflation found in~\cite{Naruko:2008sq}. In the specific cases they considered to generate a large
non-Gaussianity the non-Gaussianity was generated purely at the end of inflation and $\fnl$ is
scale independent.
In detail we see from Eqs.~(4.4) and~(4.24) in~\cite{Naruko:2008sq} that 
their formulas for $\fnl$ does not depend on $N$ or on any quantities 
evaluated at Hubble exit. This is also in contrast to the exact solution considered in section \ref{sec:exactsolution}. Our expression
for $\fnl$ in eq.~(\ref{fNLconcrete2})  depends
on the initial values 
through $p$ and $q$, and
this dependence drops out at leading order in $\alpha$.
Hence $f_{NL}$ is {\it independent} of the number of
$e$--foldings and scale independent.

\section{Conclusions}\label{sec:conclusions}

We have reviewed various models which can generate a large local non-Gaussianity. A feature shared by all of these models is that they have more than one light scalar field present during inflation. This extra degree of freedom generates an isocurvature perturbation which is at least partially converted into the primordial curvature perturbation after horizon exit of the modes which are observable today. In the curvaton and modulated reheating scenarios this conversion occurs after the end of inflation, while in the inhomogeneous end of inflation scenario this conversion occurs on the non-uniform energy density hypersurface on which inflation ends. For these three scenarios the light field which generates the primordial curvature perturbation after or at the end of inflation can be treated as a test field which does not affect the inflationary dynamics.

Our main focus has been on models in which a large non-Gaussianity is generated during inflation. This can occur even within slow-roll inflation for certain potentials and certain trajectories. We have shown, at least in the case of a separable potential, that the trajectory is required to be almost entirely along the direction of one field but that the orthogonal field must become more important towards the end of inflation and hence the inflationary trajectory must curve. In absolute terms the change to the angle of the background trajectory is small (compared to a trajectory which turns by a right angle during inflation), but in relative terms it must grow by at least an order of magnitude. This is in contrast to the previous three scenarios.

We have reviewed two-field hybrid inflation as an explicit model of a separable potential where the conditions required to generate a large non-Gaussianity can be satisfied. The conditions can be satisfied for any possible combination of positive and negative $\eta$ parameters, so the potential can be bowl shaped, a hill top or have a saddle point. The main conditions which must be satisfied is that the difference of the two $\eta$ parameters must not be too small, $\eta_{\varphi\varphi}-\eta_{\chi\chi}\sim0.1$, and the value of the $\chi$ field must be very subdominant to that of the $\varphi$ field initially (or vice versa). In general, but depending on the coupling constants between the two inflaton fields and the waterfall field, there is a change to observables at the end of inflation, due to the fact that the surface on which the waterfall field is destabilised and inflation ends might not be a surface of uniform energy density. This effect is responsible for the inhomogeneous end of inflation scenario. It is then a model dependent question whether there will be further evolution to the observables during reheating in this model, this deserves further attention.

One similarity that this hybrid inflation model has together with the quadratic curvaton scenario is that in both cases the initially subdominant (approximately isocurvature) field $\chi$ has the ratio $\delta\chi/\chi\sim\zeta_{\chi}$ approximately constant and the field fluctuations do not become more non-Gaussian with time. However the effect of this light field on the primordial curvature perturbation grows, during inflation in the hybrid scenario and before the curvaton decay in the curvaton scenario. It is this non-linear transfer between the field fluctuation and $\zeta$, described by the $\delta N$ formalism, which can generate a large non-Gaussianity. Therefore the non-Gaussianity in the hybrid scenario which we have studied is generated on super horizon scales during slow-roll inflation, in a similar way to which non-Gaussianity is generated over time in the curvaton scenario before the decay of the curvaton. The evolution of $f_{NL}$ during inflation is explicitly calculated and plotted in \cite{Byrnes:2008zy}. This conclusion is somewhat different from that in \cite{Tanaka:2010km}, and we plan to elaborate on this point in a future work. For more discussion on the distinction between non-Gaussianity generated by a non-Gaussian field perturbation, and non-Gaussianity generated by a non-linear transfer between a Gaussian field perturbation and $\zeta$ see \cite{Mulryne:2009kh}. An example where the subdominant fields fluctuations can become non-Gaussian due to a large self interaction was discussed by Bernardeau \cite{Bernardeau:2010jp}.

In order to study models where slow-roll breaks down before the end of inflation, it is clearly necessary to go beyond a formalism based on the slow-roll approximation. We have shown how this can be done in the context a separable Hubble parameter instead of a separable potential and this leads to an exact expression for (the local part of) $\fnl$ in these models. As an explicit example an exact two-field solution with an exponential potential was given. For some parameter choices this leads to a strong break down of slow roll before the end of inflation, which may give rise to a large non-Gaussianity. Further work is also required for this model to understand how the potential may be modified after the end of inflation in order that reheating occurs.

Non-Gaussianity is a topical field, in which observations have improved greatly over the last decade through both studies of the CMB and large scale structure. Observations so far have heavily focused on constraining the bispectrum non-linearity parameter $\fnl$. Currently the tightest constraint comes from the WMAP satellite, assuming the local model of non-Gaussianity this constrains the amplitude of the non-Gaussian part of the primordial curvature perturbation to be less than about one thousandth the amplitude of the Gaussian perturbation. This constraint is likely to be tightened considerably by the Planck satellite, which is currently taking data, or instead there might be a detection. A detection of $\fnl$ at this level would rule out simplest models of inflation, which are single field with a canonical kinetic term. Clearly this would be an extremely exciting result. 

However even if we are in the fortunate position of having a detection of $\fnl$ as well as improved constraints/detection of the scalar-to-tensor ratio and the spectral index, there will probably still be several viable scenarios, as detailed in this article, which for suitable parameter choices and initial conditions can match the observations. Fortunately non-Gaussianity is about much more than one number. The trispectrum (four-point function) depends on two non-linearity parameters. In general $\tau_{NL}\geq(6\fnl/5)^2$. If the current observational hints (which are not statistically significant) that $\fnl\sim40$ turn out to be true, then both the bispectrum and the trispectrum should be large enough for Planck to detect. Even if the bispectrum turns out to be much smaller, although for many models $\tau_{NL}$ is close to the lower bound, we have seen that in the model of hybrid inflation it is possible to have $\tau_{NL}\gg\fnl^2$, so the trispectrum might even be the first observational signature of non-Gaussianity. Alternatively the trispectrum through a large $g_{NL}$ might give the first observational signature, as is possible in self-interacting curvaton models or the exact solution with an exponential potential. If $\fnl$ is detected it will also be possible to either constrain or detect a scale dependence of this parameter. Although it is often assumed to be constant, this is only true for certain simple models, and for example in the two-field hybrid inflation model it generally has a significant scale dependence. We have therefore seen that non-Gaussianity is an important and powerful method of constraining and distinguishing between the many models of inflation.

\section{Acknowledgments}
The authors are extremely grateful to their collaborators with whom they have worked on numerous projects related to non-Gaussianity, which part of this review article is based on. The authors thank Takahiro Tanaka, Teruaki Suyama and Shuichiro Yokoyama
for discussions. K.Y. Choi was partly supported by the Korea Research
Foundation Grant funded by the Korean Government (KRF-2008-341-C00008)
and by the second stage of Brain Korea 21 Project in 2006.



\begin{thebibliography}{99}

\bibitem{Komatsu:2001rj}
  E.~Komatsu and D.~N.~Spergel,
  Phys.\ Rev.\  D {\bf 63} (2001) 063002
  [arXiv:astro-ph/0005036].

\bibitem{Komatsu:2010fb}
  E.~Komatsu {\it et al.},
  arXiv:1001.4538 [astro-ph.CO].


\bibitem{Komatsu:2010hc}
  E.~Komatsu,
  arXiv:1003.6097 [astro-ph.CO].

\bibitem{Liguori:2010hx}
  M.~Liguori, E.~Sefusatti, J.~R.~Fergusson and E.~P.~S.~Shellard,
  arXiv:1001.4707 [astro-ph.CO].

\bibitem{Verde:2010wp}
  L.~Verde,
  arXiv:1001.5217 [astro-ph.CO].


\bibitem{starob85}
 A.~A.~Starobinsky,
  JETP Lett.\  {\bf 42}, 152 (1985)
  [Pisma Zh.\ Eksp.\ Teor.\ Fiz.\  {\bf 42}, 124 (1985)].

\bibitem{ss1}
 M.~Sasaki   and E.~D.~Stewart,
  Prog.\ Theor.\ Phys.\  {\bf 95} (1996) 71
[arXiv:astro-ph/9507001].

\bibitem{Sasaki:1998ug}
 M.~Sasaki and T.~Tanaka,
 Prog.\ Theor.\ Phys.\  {\bf 99}, 763 (1998)
 [arXiv:gr-qc/9801017].

\bibitem{lms}
D.~H.~Lyth, K.~A.~Malik and M.~Sasaki,
JCAP {\bf 0505}, 004 (2005)
[arXiv:astro-ph/0411220].

\bibitem{Lyth:2005fi}
  D.~H.~Lyth and Y.~Rodriguez,
  Phys.\ Rev.\ Lett.\  {\bf 95} (2005) 121302
  [arXiv:astro-ph/0504045].

\bibitem{Takamizu:2010xy}
  Y.~i.~Takamizu, S.~Mukohyama, M.~Sasaki and Y.~Tanaka,
  arXiv:1004.1870 [astro-ph.CO].

\bibitem{Maldacena:2002vr}
  J.~M.~Maldacena,
  JHEP {\bf 0305} (2003) 013
  [arXiv:astro-ph/0210603].

\bibitem{Chen:2006xjb}
  X.~Chen, R.~Easther and E.~A.~Lim,
  JCAP {\bf 0706}, 023 (2007)
  [arXiv:astro-ph/0611645].


\bibitem{DBI}
  E.~Silverstein and D.~Tong,
  Phys.\ Rev.\  D {\bf 70}, 103505 (2004)
  [arXiv:hep-th/0310221];
  M.~Alishahiha, E.~Silverstein and D.~Tong,
  Phys.\ Rev.\  D {\bf 70}, 123505 (2004)
  [arXiv:hep-th/0404084];
X.~Chen, M.~X.~Huang, S.~Kachru and G.~Shiu,
arXiv:hep-th/0605045;
  M.~x.~Huang and G.~Shiu,
  Phys.\ Rev.\  D {\bf 74}, 121301 (2006)
  [arXiv:hep-th/0610235];
  D.~Langlois, S.~Renaux-Petel, D.~A.~Steer and T.~Tanaka,
  arXiv:0806.0336 [hep-th];
  F.~Arroja, S.~Mizuno and K.~Koyama,
  JCAP {\bf 0808}, 015 (2008)
  [arXiv:0806.0619 [astro-ph]];
  X.~Gao and B.~Hu,
  arXiv:0903.1920 [astro-ph.CO];
  Y.~F.~Cai and H.~Y.~Xia,
  Phys.\ Lett.\  B {\bf 677}, 226 (2009)
  [arXiv:0904.0062 [hep-th]];
  X.~Chen, B.~Hu, M.~x.~Huang, G.~Shiu and Y.~Wang,
  arXiv:0905.3494 [astro-ph.CO];
  F.~Arroja, S.~Mizuno, K.~Koyama and T.~Tanaka,
  arXiv:0905.3641 [hep-th];
  S.~Mizuno, F.~Arroja, K.~Koyama and T.~Tanaka,
  arXiv:0905.4557 [hep-th].

\bibitem{Koyama:2010xj}
  K.~Koyama,
  arXiv:1002.0600 [hep-th].

\bibitem{Chen:2010xk}
  X.~Chen,
  arXiv:1002.1416 [astro-ph.CO].

\bibitem{preheating}
  K.~Enqvist, A.~Jokinen, A.~Mazumdar, T.~Multamaki and A.~Vaihkonen,
  Phys.\ Rev.\ Lett.\  {\bf 94} (2005) 161301
  [arXiv:astro-ph/0411394];
  K.~Enqvist, A.~Jokinen, A.~Mazumdar, T.~Multamaki and A.~Vaihkonen,
  JCAP {\bf 0503} (2005) 010
  [arXiv:hep-ph/0501076];
  A.~Jokinen and A.~Mazumdar,
  JCAP {\bf 0604} (2006) 003
  [arXiv:astro-ph/0512368].

\bibitem{Lyth:2002my}
  D.~H.~Lyth, C.~Ungarelli and D.~Wands,
  Phys.\ Rev.\  D {\bf 67} (2003) 023503
  [arXiv:astro-ph/0208055].

\bibitem{Bartolo:2003jx}
  N.~Bartolo, S.~Matarrese and A.~Riotto,
  Phys.\ Rev.\  D {\bf 69} (2004) 043503
  [arXiv:hep-ph/0309033].
\bibitem{Malik:2006pm}
  K.~A.~Malik and D.~H.~Lyth,
  JCAP {\bf 0609} (2006) 008
  [arXiv:astro-ph/0604387].

\bibitem{Zaldarriaga:2003my}
  M.~Zaldarriaga,
  Phys.\ Rev.\  D {\bf 69}, 043508 (2004)
  [arXiv:astro-ph/0306006].

\bibitem{Lyth:2005qk}
  D.~H.~Lyth,
  JCAP {\bf 0511} (2005) 006
  [arXiv:astro-ph/0510443].

\bibitem{Valiviita:2009bp}
  J.~Valiviita and T.~Giannantonio,
  Phys.\ Rev.\  D {\bf 80}, 123516 (2009)
  [arXiv:0909.5190 [astro-ph.CO]].

\bibitem{Hikage:2009rt}
  C.~Hikage, D.~Munshi, A.~Heavens and P.~Coles,
  arXiv:0907.0261 [astro-ph.CO].


\bibitem{Bartolo:2010qu}
  N.~Bartolo, S.~Matarrese and A.~Riotto,
  arXiv:1001.3957 [astro-ph.CO].

\bibitem{Pitrou:2010sn}
  C.~Pitrou, J.~P.~Uzan and F.~Bernardeau,
  arXiv:1003.0481 [astro-ph.CO].

\bibitem{Wands:2007bd}
  D.~Wands,
  Lect.\ Notes Phys.\  {\bf 738} (2008) 275
  [arXiv:astro-ph/0702187].

\bibitem{Komatsu:2009kd}
  E.~Komatsu {\it et al.},
  arXiv:0902.4759 [astro-ph.CO].

\bibitem{Fergusson:2008ra}
  D.~Babich, P.~Creminelli and M.~Zaldarriaga,
  JCAP {\bf 0408} (2004) 009
  [arXiv:astro-ph/0405356]; 
  J.~R.~Fergusson and E.~P.~S.~Shellard,
  arXiv:0812.3413 [astro-ph].

\bibitem{Alabidi:2010ba}
  L.~Alabidi, K.~A.~Malik, C.~T.~Byrnes and K.~Y.~Choi,
  arXiv:1002.1700 [astro-ph.CO].

\bibitem{Mazumdar:2010sa}
  A.~Mazumdar and J.~Rocher,
  arXiv:1001.0993 [hep-ph].

\bibitem{Bernardeau:2010jp}
  F.~Bernardeau,
  arXiv:1003.2869 [astro-ph.CO].

\bibitem{Wands:2010af}
  D.~Wands,
  arXiv:1004.0818 [astro-ph.CO].

\bibitem{Tanaka:2010km}
  T.~Tanaka, T.~Suyama and S.~Yokoyama,
  arXiv:1003.5057 [astro-ph.CO].


\bibitem{Langlois:2009jp}
  D.~Langlois and L.~Sorbo,
  JCAP {\bf 0908}, 014 (2009)
  [arXiv:0906.1813 [astro-ph.CO]].

\bibitem{Barnaby:2008fk}
  N.~Barnaby and J.~M.~Cline,
  JCAP {\bf 0806}, 030 (2008)
  [arXiv:0802.3218 [hep-th]].


\bibitem{Lehners:2009qu}
  J.~L.~Lehners and P.~J.~Steinhardt,
  Phys.\ Rev.\  D {\bf 80}, 103520 (2009)
  [arXiv:0909.2558 [hep-th]].

\bibitem{Lehners:2010fy}
  J.~L.~Lehners,
  arXiv:1001.3125 [hep-th].

\bibitem{Koyama:2007if}
  K.~Koyama, S.~Mizuno, F.~Vernizzi and D.~Wands,
  JCAP {\bf 0711}, 024 (2007)
  [arXiv:0708.4321 [hep-th]].


\bibitem{Mollerach:1989hu}
  S.~Mollerach,
  Phys.\ Rev.\  D {\bf 42} (1990) 313.
\bibitem{Linde:1996gt}
  A.~D.~Linde and V.~F.~Mukhanov,
  Phys.\ Rev.\  D {\bf 56} (1997) 535
  [arXiv:astro-ph/9610219].

\bibitem{Enqvist:2001zp}
  K.~Enqvist and M.~S.~Sloth,
  Nucl.\ Phys.\  B {\bf 626} (2002) 395
  [arXiv:hep-ph/0109214].

\bibitem{Lyth:2001nq}
  D.~H.~Lyth and D.~Wands,
  Phys.\ Lett.\  B {\bf 524} (2002) 5
  [arXiv:hep-ph/0110002].
\bibitem{Moroi:2001ct}
  T.~Moroi and T.~Takahashi,
  Phys.\ Lett.\  B {\bf 522} (2001) 215
  [Erratum-ibid.\  B {\bf 539} (2002) 303]
  [arXiv:hep-ph/0110096].


\bibitem{Sasaki:2006kq}
  M.~Sasaki, J.~Valiviita and D.~Wands,
  Phys.\ Rev.\  D {\bf 74} (2006) 103003
  [arXiv:astro-ph/0607627].


\bibitem{Langlois:2008vk}
  D.~Langlois, F.~Vernizzi and D.~Wands,
  JCAP {\bf 0812} (2008) 004
  [arXiv:0809.4646 [astro-ph]].

\bibitem{Choi:2007fya}
  K.~Y.~Choi and J.~O.~Gong,
  JCAP {\bf 0706} (2007) 007
  [arXiv:0704.2939 [astro-ph]].

\bibitem{Assadullahi:2007uw}
  H.~Assadullahi, J.~Valiviita and D.~Wands,
  Phys.\ Rev.\  D {\bf 76} (2007) 103003
  [arXiv:0708.0223 [hep-ph]].
\bibitem{Huang:2008rj}
  Q.~G.~Huang,
  JCAP {\bf 0809} (2008) 017
  [arXiv:0807.1567 [hep-th]].

\bibitem{Gordon:2002gv}
  C.~Gordon and A.~Lewis,
  Phys.\ Rev.\  D {\bf 67} (2003) 123513
  [arXiv:astro-ph/0212248].
\bibitem{:2008ei}
  M.~Beltran,
  Phys.\ Rev.\  D {\bf 78} (2008) 023530
  [arXiv:0804.1097 [astro-ph]].

\bibitem{Nakayama:2009ce}
  K.~Nakayama and J.~Yokoyama,
  JCAP {\bf 1001} (2010) 010
  [arXiv:0910.0715 [astro-ph.CO]].

\bibitem{Bassett:2005xm}
  B.~A.~Bassett, S.~Tsujikawa and D.~Wands,
  Rev.\ Mod.\ Phys.\  {\bf 78}, 537 (2006)
  [arXiv:astro-ph/0507632].
  
\bibitem{Frolov:2010sz}
  A.~V.~Frolov,
  arXiv:1004.3559 [gr-qc].


\bibitem{Dvali:2003em}
  G.~Dvali, A.~Gruzinov and M.~Zaldarriaga,
  Phys.\ Rev.\  D {\bf 69} (2004) 023505
  [arXiv:astro-ph/0303591].
\bibitem{Kofman:2003nx}
  L.~Kofman,
  arXiv:astro-ph/0303614.
\bibitem{Dvali:2003ar}
  G.~Dvali, A.~Gruzinov and M.~Zaldarriaga,
  Phys.\ Rev.\  D {\bf 69} (2004) 083505
  [arXiv:astro-ph/0305548].


\bibitem{Bauer:2005cd}
  C.~W.~Bauer, M.~L.~Graesser and M.~P.~Salem,
  Phys.\ Rev.\  D {\bf 72} (2005) 023512
  [arXiv:astro-ph/0502113].

\bibitem{Suyama:2007bg}
  T.~Suyama and M.~Yamaguchi,
  Phys.\ Rev.\  D {\bf 77}, 023505 (2008)
  [arXiv:0709.2545 [astro-ph]].

\bibitem{Ichikawa:2008ne}
  K.~Ichikawa, T.~Suyama, T.~Takahashi and M.~Yamaguchi,
  Phys.\ Rev.\  D {\bf 78}, 063545 (2008)
  [arXiv:0807.3988 [astro-ph]].


\bibitem{BasteroGil:2003tj}
  M.~Bastero-Gil, V.~Di Clemente and S.~F.~King,
  Phys.\ Rev.\  D {\bf 70} (2004) 023501
  [arXiv:hep-ph/0311237].

\bibitem{Bond:2009xx}
  J.~R.~Bond, A.~V.~Frolov, Z.~Huang and L.~Kofman,
  Phys.\ Rev.\ Lett.\  {\bf 103}, 071301 (2009)
  [arXiv:0903.3407 [astro-ph.CO]].

\bibitem{Kohri:2009ac}
  K.~Kohri, D.~H.~Lyth and C.~A.~Valenzuela-Toledo,
  arXiv:0904.0793 [hep-ph].

\bibitem{felder}
  G.~N.~Felder, L.~Kofman and A.~D.~Linde,
  Phys.\ Rev.\  D {\bf 59}, 123523 (1999)
  [arXiv:hep-ph/9812289].

\bibitem{Kolb:2004jm}
  E.~W.~Kolb, A.~Riotto and A.~Vallinotto,
  Phys.\ Rev.\  D {\bf 71} (2005) 043513
  [arXiv:astro-ph/0410546].

\bibitem{Byrnes:2005th}
  C.~T.~Byrnes and D.~Wands,
  Phys.\ Rev.\  D {\bf 73}, 063509 (2006)
  [arXiv:astro-ph/0512195].

\bibitem{Matsuda:2006ee}
  T.~Matsuda,
  JCAP {\bf 0703}, 003 (2007)
  [arXiv:hep-th/0610232].

\bibitem{Battefeld:2007st}
  T.~Battefeld,
  Phys.\ Rev.\  D {\bf 77}, 063503 (2008)
  [arXiv:0710.2540 [hep-th]].

\bibitem{Byrnes:2008zz}
  C.~T.~Byrnes,
  JCAP {\bf 0901}, 011 (2009)
  [arXiv:0810.3913 [astro-ph]].


\bibitem{Huang:2009vk}
  Q.~G.~Huang,
  JCAP {\bf 0906}, 035 (2009)
  [arXiv:0904.2649 [hep-th]].

\bibitem{Bernardeau:2002jy}
  F.~Bernardeau and J.~P.~Uzan,
  Phys.\ Rev.\  D {\bf 66} (2002) 103506
  [arXiv:hep-ph/0207295].
\bibitem{Bernardeau:2002jf}
  F.~Bernardeau and J.~P.~Uzan,
  Phys.\ Rev.\  D {\bf 67} (2003) 121301
  [arXiv:astro-ph/0209330].
\bibitem{Bernardeau:2007xi}
  F.~Bernardeau and T.~Brunier,
  Phys.\ Rev.\  D {\bf 76} (2007) 043526
  [arXiv:0705.2501 [hep-ph]].

\bibitem{Salem:2005nd}
  M.~P.~Salem,
  Phys.\ Rev.\  D {\bf 72} (2005) 123516
  [arXiv:astro-ph/0511146].


\bibitem{Alabidi:2006wa}
  L.~Alabidi and D.~Lyth,
  JCAP {\bf 0608} (2006) 006
  [arXiv:astro-ph/0604569].



\bibitem{Sasaki:2008uc}
  M.~Sasaki,
  Prog.\ Theor.\ Phys.\  {\bf 120} (2008) 159
  [arXiv:0805.0974 [astro-ph]].

\bibitem{Naruko:2008sq}
  A.~Naruko and M.~Sasaki,
  Prog.\ Theor.\ Phys.\  {\bf 121}, 193 (2009)
  [arXiv:0807.0180 [astro-ph]].

\bibitem{Byrnes:2008wi}
  C.~T.~Byrnes, K.~Y.~Choi and L.~M.~H.~Hall,
  JCAP {\bf 0810} (2008) 008
  [arXiv:0807.1101 [astro-ph]].

\bibitem{Byrnes:2008zy}
  C.~T.~Byrnes, K.~Y.~Choi and L.~M.~H.~Hall,
  JCAP {\bf 0902} (2009) 017
  [arXiv:0812.0807 [astro-ph]].




\bibitem{Rigopoulos:2004gr}
  G.~I.~Rigopoulos and E.~P.~S.~Shellard,
  JCAP {\bf 0510} (2005) 006
  [arXiv:astro-ph/0405185].
\bibitem{Rigopoulos:2005xx}
  G.~I.~Rigopoulos, E.~P.~S.~Shellard and B.~J.~W.~van Tent,
  Phys.\ Rev.\  D {\bf 73} (2006) 083521
  [arXiv:astro-ph/0504508].


\bibitem{Alabidi:2005qi}
  L.~Alabidi and D.~H.~Lyth,
  JCAP {\bf 0605}, 016 (2006)
  [arXiv:astro-ph/0510441].

\bibitem{Vernizzi:2006ve}
  F.~Vernizzi and D.~Wands,
  JCAP {\bf 0605}, 019 (2006)
  [arXiv:astro-ph/0603799].

\bibitem{Rigopoulos:2005us}
  G.~I.~Rigopoulos, E.~P.~S.~Shellard and B.~J.~W.~van Tent,
  Phys.\ Rev.\  D {\bf 76}, 083512 (2007)
  [arXiv:astro-ph/0511041].

\bibitem{Yokoyama:2007uu}
  S.~Yokoyama, T.~Suyama and T.~Tanaka,
  Phys.\ Rev.\  D {\bf 77}, 083511 (2008)
  [arXiv:0705.3178 [astro-ph]];

\bibitem{Yokoyama:2007dw}
  S.~Yokoyama, T.~Suyama and T.~Tanaka,
  arXiv:0711.2920 [astro-ph].

\bibitem{Cogollo:2008bi}
  H.~R.~S.~Cogollo, Y.~Rodriguez and C.~A.~Valenzuela-Toledo,
  JCAP {\bf 0808} (2008) 029
  [arXiv:0806.1546 [astro-ph]].

\bibitem{Choi:2007su}
 K.~Y.~Choi, L.~M.~H.~Hall and C.~van de Bruck,
  JCAP {\bf 0702} (2007) 029
  [arXiv:astro-ph/0701247].
\bibitem{Battefeld:2006sz}
  T.~Battefeld and R.~Easther,
  JCAP {\bf 0703}, 020 (2007)
  [arXiv:astro-ph/0610296].


\bibitem{Seery:2006js}
  D.~Seery and J.~E.~Lidsey,
  JCAP {\bf 0701}, 008 (2007)
  [arXiv:astro-ph/0611034].

\bibitem{GarciaBellido:1995qq}
  J.~Garcia-Bellido and D.~Wands,
  Phys.\ Rev.\  D {\bf 53} (1996) 5437
  [arXiv:astro-ph/9511029].

\bibitem{Bartolo:2002vf}
  N.~Bartolo and A.~R.~Liddle,
  Phys.\ Rev.\  D {\bf 65}, 121301 (2002)
  [arXiv:astro-ph/0203076].

\bibitem{Mulryne:2009kh}
  D.~J.~Mulryne, D.~Seery and D.~Wesley,
  JCAP {\bf 1001}, 024 (2010)
  [arXiv:0909.2256 [astro-ph.CO]].

\bibitem{cline}
 N.~Barnaby and J.~M.~Cline,
 Phys.\ Rev.\ D {\bf 73}, 106012 (2006)
 [arXiv:astro-ph/0601481].


\bibitem{Mulryne:2009ci}
  D.~Mulryne, D.~Seery and D.~Wesley,
  arXiv:0911.3550 [astro-ph.CO].


\bibitem{Byrnes:2009qy}
  C.~T.~Byrnes and G.~Tasinato,
  JCAP {\bf 0908}, 016 (2009)
  [arXiv:0906.0767 [astro-ph.CO]].

\bibitem{SB}
  D.~S.~Salopek and J.~R.~Bond,
  Phys.\ Rev.\  D {\bf 42} (1990) 3936.
  
\bibitem{kinney}
  W.~H.~Kinney,
  Phys.\ Rev.\  D {\bf 56} (1997) 2002
  [arXiv:hep-ph/9702427].
    
\bibitem{Battefeld:2009ym}
  D.~Battefeld and T.~Battefeld,
  JCAP {\bf 0911}, 010 (2009)
  [arXiv:0908.4269 [hep-th]].
\bibitem{Choi:2008et}
  K.~Y.~Choi, J.~O.~Gong and D.~Jeong,
  JCAP {\bf 0902}, 032 (2009)
  [arXiv:0810.2299 [hep-ph]].


\bibitem{Sasaki:2007ay}
  M.~Sasaki,
  Class.\ Quant.\ Grav.\  {\bf 24}, 2433 (2007)
  [arXiv:astro-ph/0702182].

\bibitem{kahlerinfl}
  J.~P.~Conlon and F.~Quevedo,
  JHEP {\bf 0601} (2006) 146
  [arXiv:hep-th/0509012];
  J.~R.~Bond, L.~Kofman, S.~Prokushkin and P.~M.~Vaudrevange,
  Phys.\ Rev.\  D {\bf 75} (2007) 123511
  [arXiv:hep-th/0612197]; 
  A.~Misra and P.~Shukla,
  Nucl.\ Phys.\  B {\bf 800} (2008) 384
  [arXiv:0712.1260 [hep-th]];
  M.~Badziak and M.~Olechowski,
  JCAP {\bf 0807} (2008) 021
  [arXiv:0802.1014 [hep-th]];
  E.~Palti, G.~Tasinato and J.~Ward,
  JHEP {\bf 0806} (2008) 084
  [arXiv:0804.1248 [hep-th]];
  H.~X.~Yang and H.~L.~Ma,
  JCAP {\bf 0808} (2008) 024
  [arXiv:0804.3653 [hep-th]];
  J.~J.~Blanco-Pillado, D.~Buck, E.~J.~Copeland, M.~Gomez-Reino and N.~J.~Nunes,
  arXiv:0906.3711 [hep-th];
  N.~Barnaby, J.~R.~Bond, Z.~Huang and L.~Kofman,
  arXiv:0909.0503 [hep-th].

\bibitem{Komatsu:2008hk}
  E.~Komatsu {\it et al.}  [WMAP Collaboration],
  arXiv:0803.0547 [astro-ph].

\bibitem{Byrnes:2006vq}
  C.~T.~Byrnes, M.~Sasaki and D.~Wands,
  Phys.\ Rev.\  D {\bf 74}, 123519 (2006)
  [arXiv:astro-ph/0611075].

\bibitem{Smidt:2010sv}
  J.~Smidt, A.~Amblard, A.~Cooray, A.~Heavens, D.~Munshi and P.~Serra,
  arXiv:1001.5026 [astro-ph.CO].

\bibitem{Desjacques:2009jb}
  V.~Desjacques and U.~Seljak,
  arXiv:0907.2257 [astro-ph.CO].

\bibitem{Vielva:2009jz}
  P.~Vielva and J.~L.~Sanz,
  arXiv:0910.3196 [astro-ph.CO].

\bibitem{Kogo:2006kh}
  T.~Okamoto and W.~Hu,
  Phys.\ Rev.\  D {\bf 66}, 063008 (2002)
  [arXiv:astro-ph/0206155];
  N.~Kogo and E.~Komatsu,
  Phys.\ Rev.\  D {\bf 73}, 083007 (2006)
  [arXiv:astro-ph/0602099];
  A.~Cooray, C.~Li and A.~Melchiorri,
  Phys.\ Rev.\  D {\bf 77}, 103506 (2008)
  [arXiv:0801.3463 [astro-ph]];
  D.~Jeong and E.~Komatsu,
  arXiv:0904.0497 [astro-ph.CO].

\bibitem{Smidt:2010ra}
  J.~Smidt, A.~Amblard, C.~T.~Byrnes, A.~Cooray and D.~Munshi,
  arXiv:1004.1409 [astro-ph.CO].







\bibitem{Creminelli:2008es}
  P.~Creminelli, S.~Dubovsky, A.~Nicolis, L.~Senatore and M.~Zaldarriaga,
  arXiv:0802.1067 [hep-th].
  
\bibitem{Rodriguez:2008hy}
  Y.~Rodriguez and C.~A.~Valenzuela-Toledo,
  arXiv:0811.4092 [astro-ph].

\bibitem{Enqvist:2005pg}
  K.~Enqvist and S.~Nurmi,
  JCAP {\bf 0510}, 013 (2005)
  [arXiv:astro-ph/0508573].

\bibitem{Huang:2008bg}
  Q.~G.~Huang and Y.~Wang,
  JCAP {\bf 0809}, 025 (2008)
  [arXiv:0808.1168 [hep-th]].

\bibitem{Huang:2008zj}
  Q.~G.~Huang,
  JCAP {\bf 0811}, 005 (2008)
  [arXiv:0808.1793 [hep-th]].

\bibitem{Enqvist:2008gk}
  K.~Enqvist and T.~Takahashi,
  JCAP {\bf 0809}, 012 (2008)
  [arXiv:0807.3069 [astro-ph]].

\bibitem{Enqvist:2009ww}
  K.~Enqvist, S.~Nurmi, O.~Taanila and T.~Takahashi,
  arXiv:0912.4657 [astro-ph.CO].


\bibitem{Byrnes:2009pe}
  C.~T.~Byrnes, S.~Nurmi, G.~Tasinato and D.~Wands,
  arXiv:0911.2780 [Unknown].

\bibitem{Sefusatti:2009xu}
  E.~Sefusatti, M.~Liguori, A.~P.~S.~Yadav, M.~G.~Jackson and E.~Pajer,
  arXiv:0906.0232 [astro-ph.CO].
 
\bibitem{LoVerde:2007ri}
  M.~LoVerde, A.~Miller, S.~Shandera and L.~Verde,
  JCAP {\bf 0804}, 014 (2008)
  [arXiv:0711.4126 [astro-ph]].
  
\bibitem{Chen:2005fe}
  X.~Chen,
  Phys.\ Rev.\  D {\bf 72}, 123518 (2005)
  [arXiv:astro-ph/0507053].

\bibitem{Khoury:2008wj}
  J.~Khoury and F.~Piazza,
  JCAP {\bf 0907}, 026 (2009)
  [arXiv:0811.3633 [hep-th]].

\end{thebibliography}
\end{document}